\documentclass[a4paper,12pt]{article}

\usepackage{lmodern}
\usepackage[T1]{fontenc}
\usepackage[utf8]{inputenc}
\usepackage{textcomp}
\usepackage[french,english]{babel}
\usepackage{amssymb,amsmath,latexsym}
\usepackage{bbm}
\usepackage{natbib}
\usepackage{graphicx}
\usepackage{booktabs}
\usepackage{multirow}
\usepackage{colortbl}
\usepackage{rotating}
\usepackage{hyperref}
\usepackage{verbatim}

\newtheorem{theorem}{Theorem}
\newcommand{\ud}{\,\mathrm{d}}
\providecommand{\keywords}[1]{\textbf{\textit{Keywords:}} #1}

\allowdisplaybreaks

\graphicspath{{graphics/}}

\begin{document}
\newpage
\setcounter{page}{1}

\title{Managing Volatility Risk: An Application of Karhunen-Loève Decomposition and Filtered Historical Simulation}
\author{Jinglun Yao, Sabine Laurent, Brice Bénaben}
\date{\today}
\maketitle

\begin{abstract}
\noindent
Implied volatilities form a well-known structure of smile or surface which accommodates the Bachelier model and observed market prices of interest rate options. For the swaptions that we study, three parameters are taken into account for indexing the implied volatilities and form a ``volatility cube'': strike (or moneyness), time to maturity of the option contract, duration of the underlying swap contract. It should be noted that the implied volatility structure changes across time, which makes it important to study its dynamics in order to well manage the volatility risk. As volatilities are correlated across the cube, it is preferable to decompose the dynamics on orthogonal principal components, which is the idea of Karhunen-Loève decomposition that we have adopted in the article. The projections on principal components are investigated by Filtered Historical Simulation in order to predict the Value at Risk (VaR), which is then examined by standard tests and non-arbitrage condition to ensure its appropriateness.
\end{abstract}


\newpage
\section{Introduction}
Since the liquidity trap of the eurozone and negative interest rates associated with it, Bachelier model (\cite{bachelier1900theorie}) has become more appropriate than Black-Scholes model (\cite{black1973pricing}) for interest rate products because of its ability to handle negative interest rates. While volatility of the underlying asset in the most basic Bachelier model is constant, in practice we should calibrate volatilities for different derivative products (e.g. call options with different strikes) at time $t$, leading to what we call ``volatility smile'' or ``volatility surface''. It is preferable and heuristic to assume the smile to be static and invariant across time $t$, which is the case when practitioners adopt the so-called ``sticky moneyness'' or ``sticky strike'' rule. Unfortunately, however, according to \cite{rosenberg2000implied} and \cite{cont2002dynamics}, we observe in reality rather continuous change in a smile. This dynamics of volatility smiles or surfaces is important for risk management since it would affect the volatility parameter in the Bachelier pricing function, resulting in the change of values of interest rate derivatives. The bank's portfolio is thus influenced and suffers from Value at Risk (VaR). More particularly, this is what we call ``Vega Risk'' in the terminology of risk management since the risk factor is volatility. In our study, we are particularly interested in swaptions, which have 3 parameters for Bachelier implied volatility $\sigma^B_t$: strike (or moneyness), time to maturity of the option contract (hereafter ``expiry''), duration of the underlying interest rate swap contract (hereafter ``tenor''). \\

Despite the obvious importance of modeling the dynamics, it is not easy to accomplish this task. The dynamics of smiles concerns not only at-the-money (ATM) volatilities but also non-ATM ones. For the series of ATM volatilities, we might well model it using traditional time series models. Yet it is not appropriate to model non-ATM volatilities separately from the ATM ones since there exists clearly a correlation between ATM and non-ATM volatilities. The smile, though not a rigid body with only three degrees of freedom, is not a ``soft body'' with infinitely many degrees of freedom, either. The stickiness of volatilities across different strikes requires the study of smile dynamics in a holistic fashion. Some major moves are more dominant than others and it is reasonable to use them as a concise description of the dynamics. For example, parallel shift of the smile is a common phenomenon. Nonetheless, these major moves are not known a priori, which should be estimated from data instead. \\

It should be noted that we have resisted the temptation of using a local volatility model (\cite{dupire1997pricing}) or a stochastic volatility model (e.g. SABR model in \cite{hagan2002managing}). These models, by introducing an infinitesimal description of volatility, adds more degrees of freedom for the volatility and explains the empirical derivation from Bachelier model with constant volatility. Implied volatility, by contrast, is a state variable which accommodates Bachelier model and market prices. However, similar to the dynamics of implied volatility smiles, we cannot assume the parameters in local or stochastic volatility models to be constant. In fact, as \cite{hagan2002managing} remarked, the dynamics of the market smile predicted by local volatility models is opposite of observed market behavior: when the price of the underlying decreases, local volatility models predict that the smile shifts to higher prices; when the price decreases, these models predict that the smile shifts to lower prices. In reality, asset prices and market smiles move in the same direction. In consequence, frequent recalibration is needed in order to ensure the correctness of the local volatility model and model parameters bear a dynamics. This is also the case for SABR model in practice, even if asset prices and market smiles move in the same direction. The dynamics of model parameters in a local or stochastic volatility model is of course a \textit{reflection} of the smile dynamics. But it should be noted that the relationship between them is not clear-cut: a calibration method is needed for estimating model parameters since there are less parameters in say, SABR model, than the degree of freedom of volatility smile. This implies that we should establish an objective function of model parameters to optimize. Yet it is difficult to tell which calibration method is ``best '' or ``better'' since none of these methods would reproduce the empirically observed implied volatilities. What's more, even if we admit a calibration method, whether it is appropriate to use the time series of calibrated parameters for risk management remains to be a question. Because ``real'' realized values are needed in order to evaluate the performance of VaR prediction, yet the ``real'' values of model parameters are not directly (or indirectly) observable in the markets.\\

A second and deeper reason for using a model-free approach (in the sense that we do not use a local or stochastic volatility model, but we have of course used the Bachelier model) is explained by \cite{cont2002dynamics}: option markets have become increasingly autonomous and option prices are driven, in addition to movements in the underlying asset, also by internal supply and demand in the options market. For example, \cite{bakshi2000call} has documented evidence of violations of quantitative dynamical relations between options and their underlying. In fact, practitioners would resort to supply-demand equilibrium and implied volatilities instead of stochastic volatility models for liquid assets. This is the case for vanilla swaptions in our studies. Despite these arguments favoring a model-free approach, it should be remarked that stochastic volatility models provide an essential way of evaluating price and managing risk for less liquid assets which are out of the scope of this article. \\

Facing the challenges of modeling smile dynamics, we have adopted the Karhunen-Loève decomposition which can be seen as a generalized version of Principal Component Analysis conceived for functions. The method was proposed in \cite{loeve1978probability} and adopted by \cite{cont2002dynamics} for studying volatility surface dynamics of stock index options. We have extended the usage of this method to swaptions and explored the implications for risk management. Since swaptions have three parameters (moneyness, expiry, tenor) for implied volatilities, it would be interesting to explore the dynamics for each dimension. Moreover, Karhunen-Loève decomposition can be applied to multivariate functions, making it possible to explore the dynamics of surfaces. In this way, the dynamics of implied volatility smiles or surfaces can be characterized by several principal components and the time series of projections on these components. The time series of projected values can then be studied by Historical Simulation or Filtered Historical Simulation in order to evaluate VaR. Historical Simulation has been overwhelmingly popular in assessing VaR in recent years because of its non-parametric approach. And Filtered Historical Simulation, based on Historical Simulation and proposed by \cite{barone1999var}, has overcome the problem of volatility clustering illustrated in \cite{cont2001empirical}.\\

The article is organized in the following way: Section~\ref{sec:KL_theory} gives an introduction to the theory of Karhunen-Loève decomposition, which is then applied to implied volatility smiles and surfaces of swaptions in section~\ref{sec:KL_app}. Empirical evidence and intuitions for different dimensions are also explored. Section~\ref{sec:FHS} studies the properties of projected time series and uses Filtered Historical Simulation for evaluating VaR of volatility increments. Note that the most important (and perhaps the only) hypothesis of derivative pricing is non-arbitrage. Sections~\ref{sec:arbitrage} checks if the estimated VaR violates this fundamental assumption. Moreover, Section~\ref{sec:backtest} checks if the estimated VaR meets some standard backtesting criteria. Section~\ref{sec:conclusion} concludes the article. 


\section{Karhunen-Loève Decomposition: Theoretical Background}
\label{sec:KL_theory}
The Karhunen-Loève decomposition is a generalization of Principal Component Analysis conceived for functions. Let $D\subset \mathbb{R}^n$ be a bounded domain. For example, in the case where we want to study the dynamics of smiles (as a function of moneyness), $D$ could be $[m_{min}, m_{max}]$, where $m_{min}$ and $m_{max}$ are respectively the smallest and largest moneyness. Then we can define the integral operator $K$ on $L^2(D)$, $K: u \mapsto Ku$ for $u \in L^2(D)$, by:

\begin{equation}
[Ku](x)=\int_D k(x,y)u(y)\ud y
\end{equation}

It can be shown that if $k:D\times D \to \mathbb{R}$ is a Hilbert-Schmidt kernel, i.e.

\begin{equation}
\int_D\int_D |k(x,y)|^2 \ud x \ud y<\infty,
\end{equation}

then K is a compact operator. Furthermore, if $k(x, y)=k(y, x) \text{ }\forall x, y\in D$, then $K$ is a self-adjoint operator. More particularly, $K$ and $k$ are defined in the way presented below for Karhunen-Loève decomposition so that they satisfy these properties. \\

In order to get a series of smiles, we add another parameter $\omega\in \Omega$ to $u$, i.e., $u: \Omega \times D \to \mathbb{R}$. We shall not distinguish between $\omega$ and $t$ for notation since in data, each $t$ represents a realization of $\omega$. Intuitively, for each $\omega$ fixed, $u(\omega, \cdot)$ is for example a volatility smile (more precisely the log-return of the smile, for reason explained below), while for each $x$ fixed, $u(\cdot, x)$ can be seen as a random variable. So we can calculate the covariance between $u(\cdot, x)$ and $u(\cdot, y)$ and define $k(x, y)$ as follows:

\begin{equation}
k(x,y)=\text{cov}(u(\cdot, x), u(\cdot, y))
\end{equation}

Then $K$ is a compact and self-adjoint operator. Remark that the kernel function $k$ is analogous to the covariance matrix of a random vector in a conventional Principal Component Analysis. It can also be shown that $K$ is positive, so according to Mercer's theorem (\cite{mercer1909functions}), $k(x,y)$ can be represented by:

\begin{equation}
k(x,y)=\sum_i \lambda_i e_i(x) e_i(y)
\end{equation}

where $\{\lambda_i\}_i$ and $\{e_i\}_i$ are eigenvalues and eigenvectors of $K$, i.e. 

\begin{equation}
K(e_i)=\int_D k(\cdot,y)e_i(y)\ud y=\lambda_i e_i
\label{eqn:fredholm}
\end{equation}

Without loss of generality, assume that $\lambda_1\geq \lambda_2 \geq \cdots \geq 0$. Let $u_i(\omega)=\langle u(\omega, \cdot), e_i \rangle$ be the projection of a smile or surface $u(\omega, \cdot)$ on the eigenfunction $e_i$. Then

\begin{equation}
u(\omega, \cdot)=\sum_i u_i(\omega) e_i(\cdot) \quad \text{and} \quad u_i(\omega)=\int_D u(\omega, x) e_i(x) \ud x
\label{eqn:decomposition}
\end{equation}

Equation~\ref{eqn:decomposition} is what we call Karhunen-Loève decomposition. It is important to note that if $u(\cdot, x)$ is a centered variable for each $x$, then $u_i(\cdot)$ is also centered. To see this, it suffices to take expectation on Equation~\ref{eqn:decomposition}. Moreover, $\{u_i(\cdot)\}_i$ are mutually uncorrelated, since

\begin{eqnarray}
\mathbb{E}(u_i u_j) &=& \mathbb{E}[\int_D u(\cdot, x)e_i(x) \ud x \int_D u(\cdot, y) e_j(y) \ud y]\\
&=& \mathbb{E}[\int_D\int_D u(\cdot, x)u(\cdot, y)e_i(x)e_j(y) \ud x \ud y]\\
&=& \int_D\int_D\mathbb{E}[u(\cdot, x)u(\cdot, y)]e_i(x)e_j(y)\ud x \ud y\\
&=& \int_D\int_D k(x,y)e_i(x)e_j(y)\ud x \ud y\\
&=& \int_D [Ke_i](y)e_j(y) \ud y\\
&=& \langle Ke_i, e_j \rangle\\
&=& \langle \lambda_i e_i, e_j \rangle\\
&=& \lambda_i \delta_{ij}
\end{eqnarray}

where $\delta_{ij}$ is Kronecker's symbol. This can help us study separately the properties of projection times series without worrying about the correlations between them. This also explains the following notation often used in practice which is equivalent to Equation~\ref{eqn:decomposition}:

\begin{equation}
u(\omega, x)=\sum_i \sqrt{\lambda_i}\xi_i(\omega)e_i(x)
\label{eqn:decomposition_2}
\end{equation}

where $\xi_i$ are centered mutually uncorrelated random variables with unit variance and are given by:

\begin{equation}
\xi_i(w)=\frac{1}{\sqrt{\lambda_i}}\int_D u(\omega, x)e_i(x) \ud x
\end{equation}

We shall stick with this notation in the following sections.\\

How to solve the problem numerically? Suppose we have observed a random field $\{u(t, \cdot)\}_t$, we can first of all use the empirical covariance to estimate $k(x,y)$, i.e.

\begin{eqnarray}
\hat{k}(x,y)&=&\frac{1}{T}\sum_{t=1}^T [u(t, x)-\bar{u}(x)][u(t, y)-\bar{u}(y)]\\
&=&\frac{1}{T}\sum_{t=1}^T [u(t, x)-\frac{1}{T}\sum_{t=1}^T u(t,x)][u(t, y)-\frac{1}{T}\sum_{t=1}^T u(t,y)]
\end{eqnarray}

For solving the problem of eigenvalues and eigenfunctions in Equation~\ref{eqn:fredholm}, we can reduce it to a finite-dimensional problem by using a Galerkin scheme, i.e. using a linear combination of basis functions for approximating each eigenfunction. For example, we can choose Legendre functions for the basis functions. Approximately, we have:

\begin{equation}
e_i(x)=\sum_{n=1}^N d^{(i)}_n \phi_n(x)=\Phi^T(x)D^{(i)}
\label{eqn:galerkin}
\end{equation}

where $\{\phi_n\}_n$ are basis functions, $\Phi(x)$ is the vector of basis functions, $D^{(i)}$ is the vector of coefficients of shape $N\times 1$ to be estimated. Combining Equation~\ref{eqn:galerkin} and Equation~\ref{eqn:fredholm}, we get:

\begin{equation}
\sum_{n=1}^{N}d^{(i)}_n\int_D k(x, y) \phi_n(y) \ud y=\lambda_i \sum_{n=1}^{N}d^{(i)}_n\phi_n(x)
\end{equation}

Multiplying both sides by $\phi_m(x)$ and integrating on $D$, we get:

\begin{equation}
\sum_{n=1}^{N}d^{(i)}_n\int_D \int_D k(x, y) \phi_m(x) \phi_n(y) \ud x \ud y=\lambda_i \sum_{n=1}^{N}d^{(i)}_n\int_D \phi_m(x) \phi_n(x) \ud x
\end{equation}

Or

\begin{equation}
AD^{(i)}=\lambda_i BD^{(i)}
\label{eqn:num_pb}
\end{equation}
where A and B are positive symmetric and defined in the following way:

\begin{eqnarray}
A_{mn}&=&\int_D \int_D k(x, y)\phi_m(x) \phi_n(y) \ud x \ud y\\
B_{mn}&=&\int_D \phi_m(x) \phi_n(x) \ud x 
\end{eqnarray}

Numerically solving the problem~\ref{eqn:num_pb} will yield the eigenvalues and eigenfunctions of operator $K$.


\section{Karhunen-Loève Decomposition: Application on Swaption Implied Volatilities}
\label{sec:KL_app}

The structure of implied volatilities is important for accommodating Bachelier model and market price of options. For vanilla swaptions, in addition to the two parameters (strike or moneyness, time to maturity) of any European option, the duration of the underlying swap contract (``tenor'') is also a crucial parameter. In the article, moneyness is defined as 

\begin{equation}
moneyness=strike - forward\text{ }rate
\label{eqn:moneyness}
\end{equation}

Keeping expiry and tenor fixed, we can study the dynamics of moneyness-indexed smile using Karhunen-Loève decomposition. Studying the dynamics of the whole smile is important since it helps us to manage volatility risk related to skewness and convexity of the smile, in addition to the evolution of ATM volatilities which is the most important but not sufficient to fully characterize the smile dynamics. As the bank has not only ATM options but also non-ATM options, managing volatility risk related to non-ATM volatilities is crucial and requires the study of dynamics of the whole smile. The same is true for expiry-indexed smiles or tenor-indexed smiles, keeping the other two parameters fixed. It should be noted that implied volatilities $\{I(\omega, x)\}_{\omega\in\Omega}$ is obviously not centered, so we take the log-return of volatilities in order to apply the Karhunen-Loève decomposition, i.e. 

\begin{equation}
u(t, x)=log(I(t, x))-log(I(t-1, x))
\label{eqn:return}
\end{equation}

The data we use are implied volatilities for USD dollars, from 2007 to 2017. \\

\begin{figure}[htbp]
\centering
\includegraphics[width=0.8\textwidth]{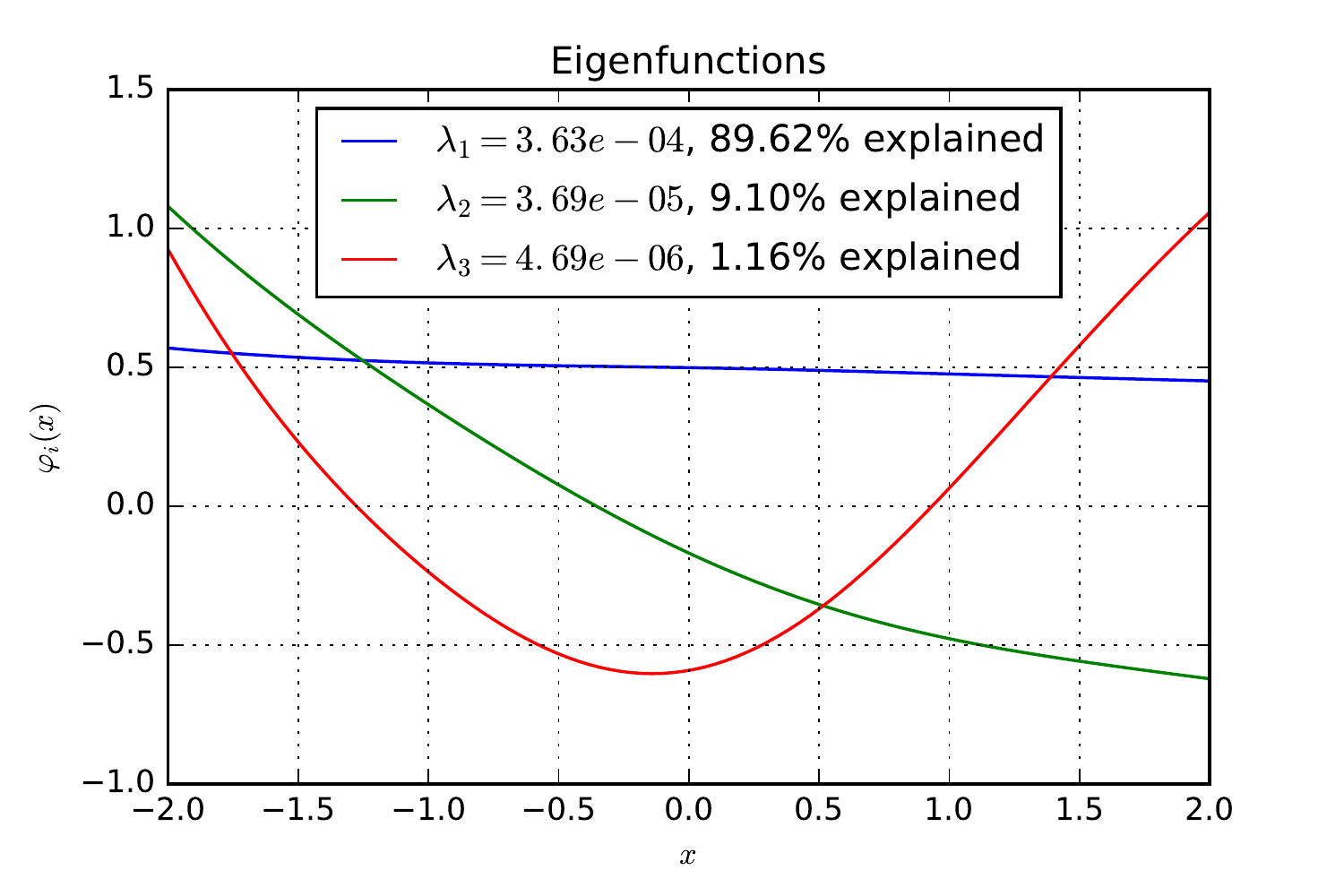}
\caption{First three eigenfunctions and eigenvalues of Karhunen-Loève decomposition for moneyness-indexed smile log-return. expiry=10Y, tenor=10Y, currency=USD. The x-axis is moneyness and has been multiplied by $100$, i.e. $x=2$ means $strike=forward\text{ }rate+2\%$}
\label{fig:eigen_moneyness_indexed_smile}
\end{figure}

\begin{figure}[htbp]
\centering
\includegraphics[width=0.8\textwidth]{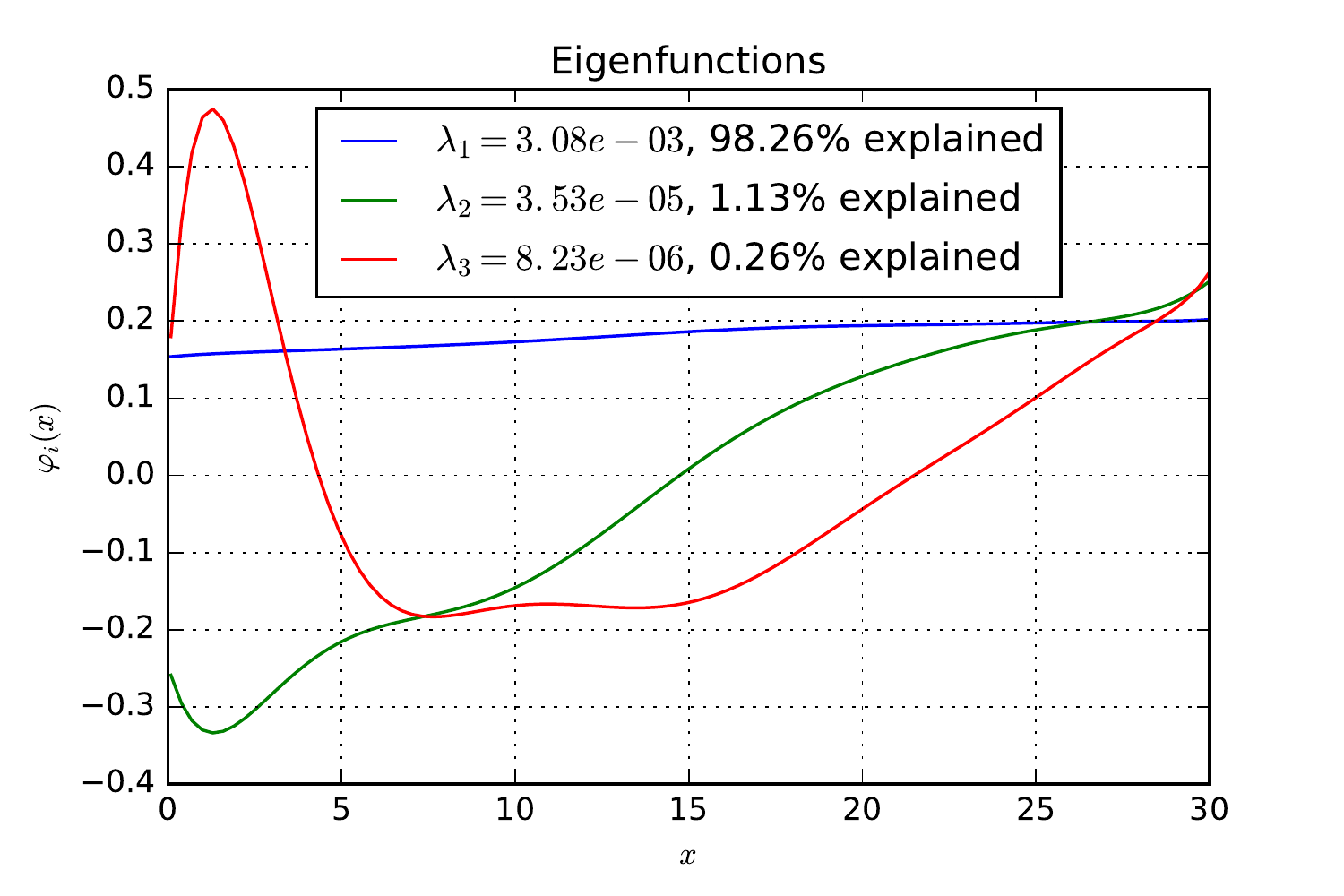}
\caption{First three eigenfunctions and eigenvalues of Karhunen-Loève decomposition for tenor-indexed smile log-return. moneyness=0, expiry=10Y, currency=USD. The unit for x-axis is year.}
\label{fig:eigen_tenor_indexed_smile}
\end{figure}

Figure~\ref{fig:eigen_moneyness_indexed_smile} illustrates the result of Karhunen-Loève decomposition for moneyness-indexed smile of swaption with expiry=10Y and tenor=10Y (hereafter ``10Y 10Y swaption''). Remark that the first eigenfunction, which can be interpreted as parallel shift of the smile, explains $89.62\%$ of the dynamics. The second eigenfunction explains most of the remaining dynamics and can be interpreted as skew change (rotation). The third eigenfunction, whose influence is more debatable, can be understood as the change of convexity. In fact, the same understanding can be gained in Figure~\ref{fig:eigen_tenor_indexed_smile} for tenor-indexed smile. \\

\begin{figure}[htbp]
\centering
\includegraphics[width=0.8\textwidth]{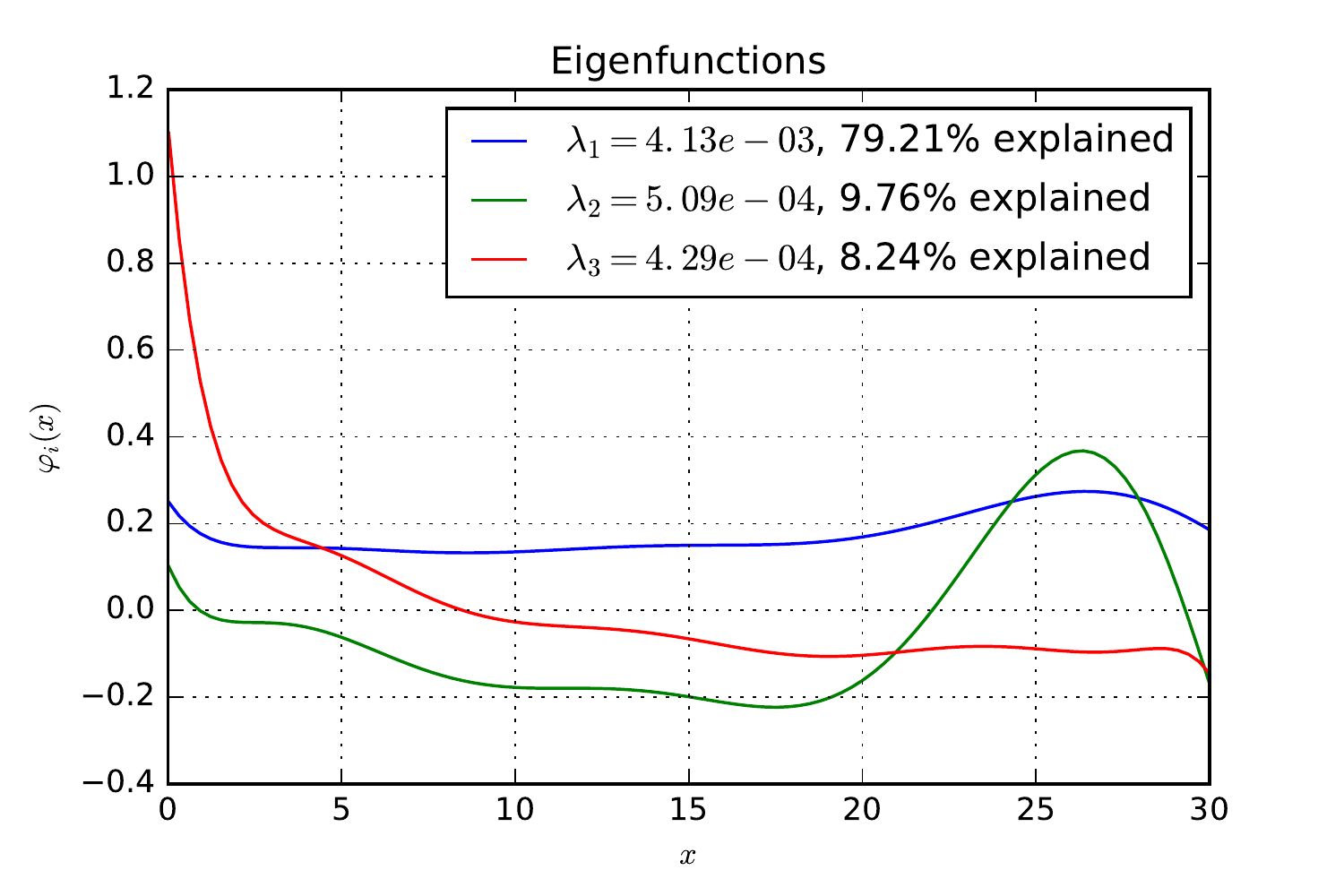}
\caption{First three eigenfunctions and eigenvalues of Karhunen-Loève decomposition for expiry-indexed smile log-return. moneyness=0, tenor=10Y, currency=USD. The unit for x-axis is year.}
\label{fig:eigen_expiry_indexed_smile}
\end{figure}

These intuitive interpretations are, however, no longer valid for expiry-indexed smile. Figure~\ref{fig:eigen_expiry_indexed_smile} illustrates the eigenfunctions and eigenvalues for log-return of expiry-indexed smile. While the first eigenfunction can still be interpreted as parallel shift, the second one, which accounts for nearly $10\%$ of the dynamics, has hardly any interpretation. This difference between expiry-indexed smile and tenor-indexed smile is consistent with practitioners' experience that tenor-indexed smile is more ``rigid'' than expiry-indexed smile. Because when tenor is fixed, different expiries would incorporate different levels of uncertainty in the option contract. The implied volatilities across different expiries are loosely related and exhibit less regularities.\\

\begin{figure}[htbp]
\centering
\includegraphics[width=0.6\textwidth]{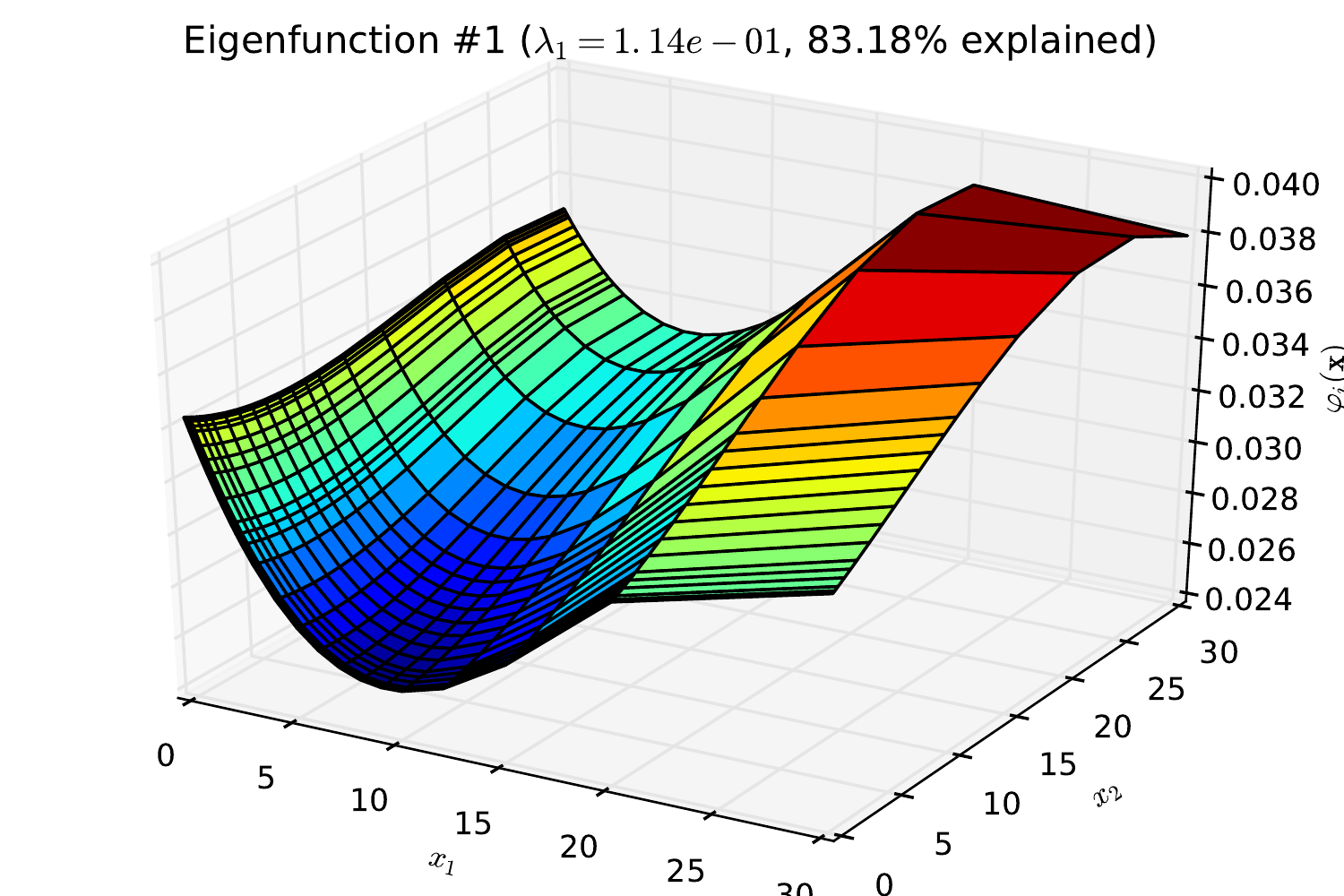}
\includegraphics[width=0.6\textwidth]{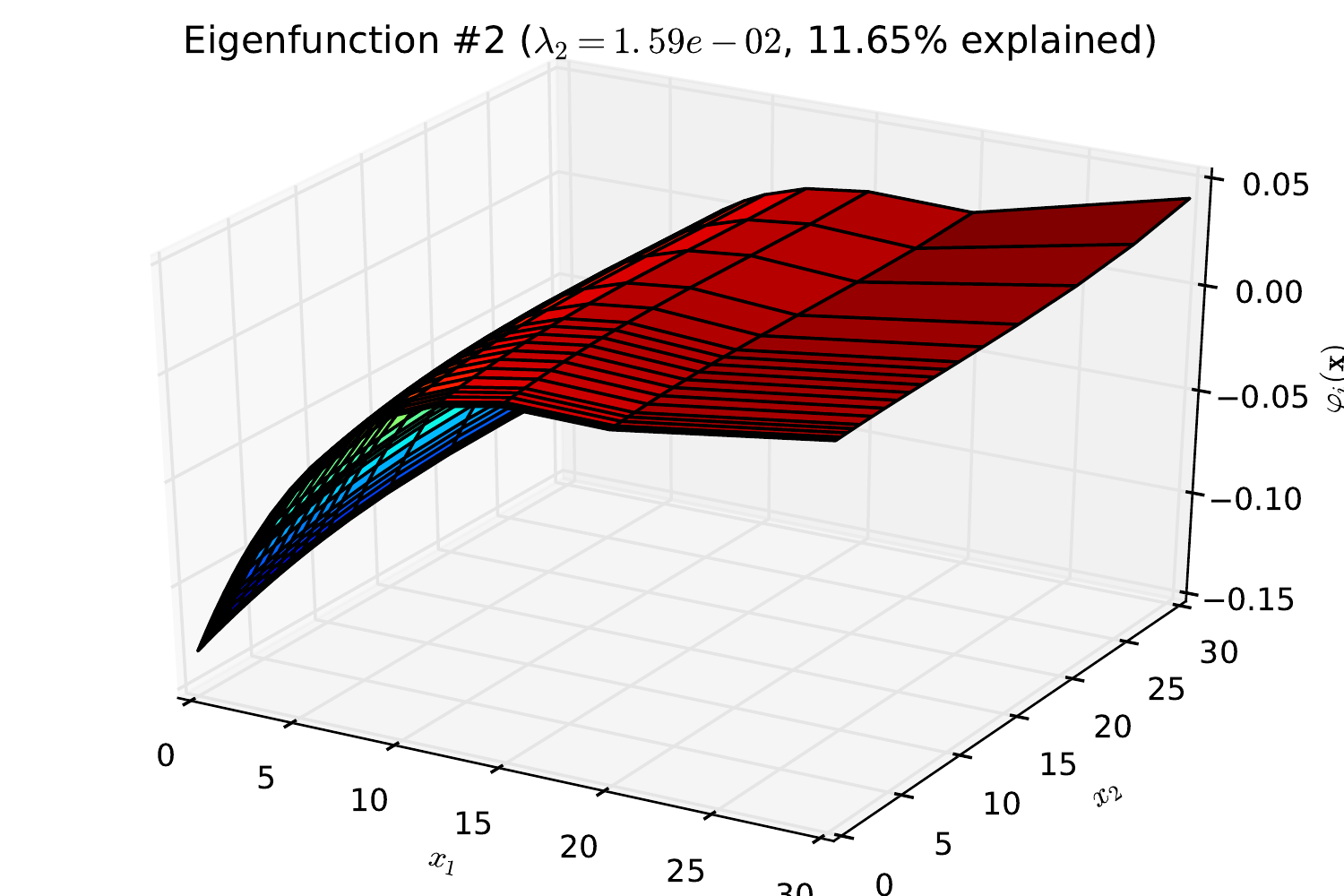}
\includegraphics[width=0.6\textwidth]{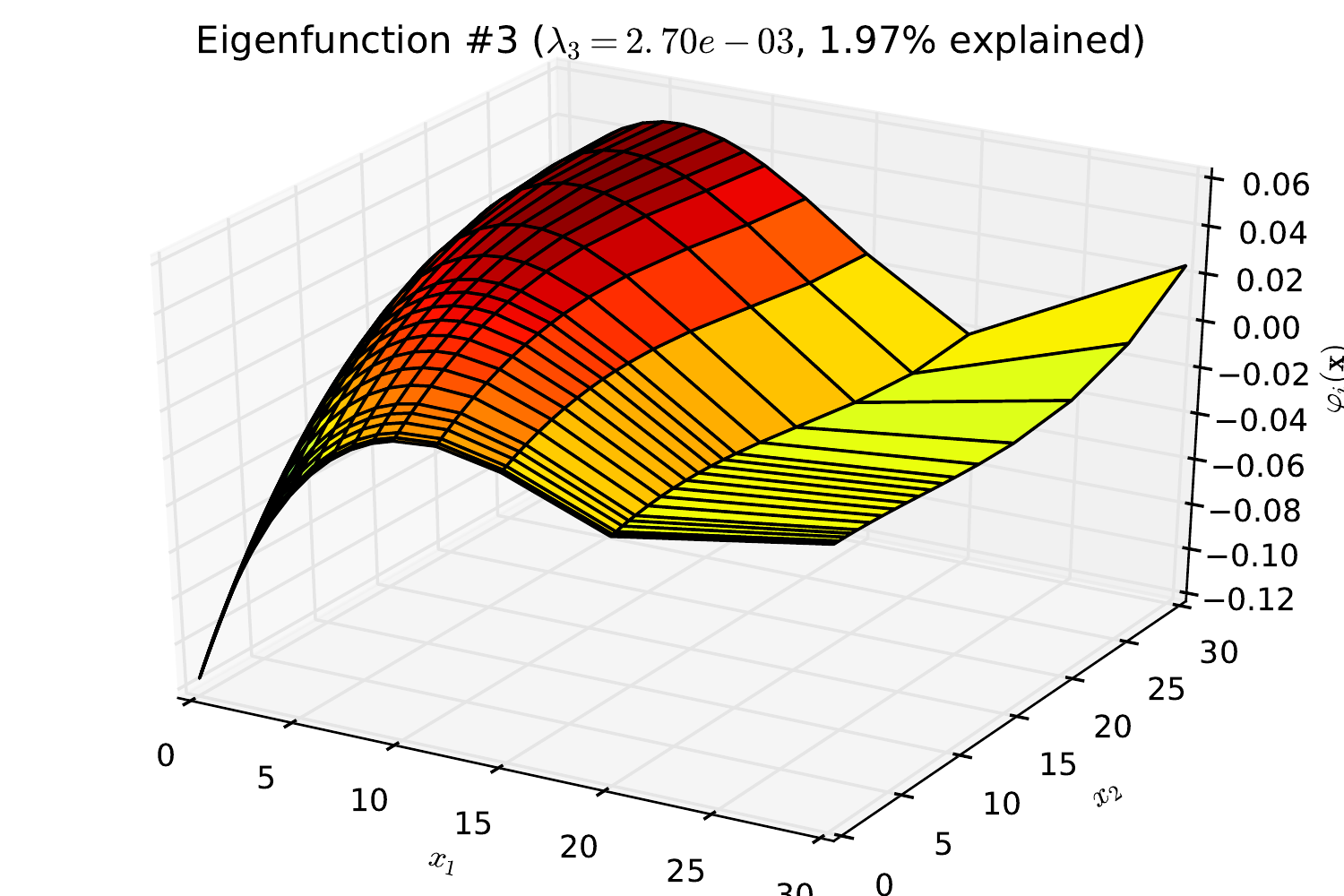}
\caption{First three eigenfunctions and eigenvalues of Karhunen-Loève decomposition for expiry-and-tenor-indexed surface log-return. moneyness=0, currency=USD. $x_1$ is expiry and $x_2$ is tenor.}
\label{fig:eigen_expiry_and_tenor_indexed_surface}
\end{figure}

Karhunen-Loève decomposition can also be applied on series of \textit{multivariate} functions. Figure~\ref{fig:eigen_expiry_and_tenor_indexed_surface} illustrates the first three eigenfunctions for log-return of expiry-and-tenor-indexed volatility surfaces. The first eigenfunction, even though not quite flat, is always positive and can be interpreted as parallel shift. The second eigenfunction reflects the rotation across the expiry since it monotonously crosses 0, but parallel shift across the tenor. One important remark is that marginal functions of the two-dimensional eigenfunctions in Figure~\ref{fig:eigen_expiry_and_tenor_indexed_surface} do not correspond exactly to the one-dimensional eigenfunctions in Figure~\ref{fig:eigen_tenor_indexed_smile} and \ref{fig:eigen_expiry_indexed_smile}. This should be due to the fact that some volatilities on the volatility surface are significantly correlated. For example, the implied volatility of 10Y 10Y swaption should be intimately linked to that of 9Y 11Y swaption because of the existence of Bermudan swaptions, which are the most actively traded exotic swaptions. For a Bermudan swaption contract which will end in 20 years, the holder of the contract may be able to exercise the contract in 9 years or 10 years. So the implied volatilities of 10Y 10Y swaption is in this way linked to the implied volatilities of 9Y 11Y swaption. It should also be noted that in Figure~\ref{fig:eigen_expiry_indexed_smile} and \ref{fig:eigen_tenor_indexed_smile}, eigenvalues for expiry-indexed smiles are larger than those for tenor-indexed smiles, given the same domain of definition $[0,30]$. So the movements related to expiry should be more significant than those related to tenor in Figure~\ref{fig:eigen_expiry_and_tenor_indexed_surface}. This might explain the second eigenfunction in Figure~\ref{fig:eigen_expiry_and_tenor_indexed_surface} in which the marginal function is approximately invariant for a fixed expiry.

\begin{figure}[htbp]
\centering
\includegraphics[width=0.6\textwidth]{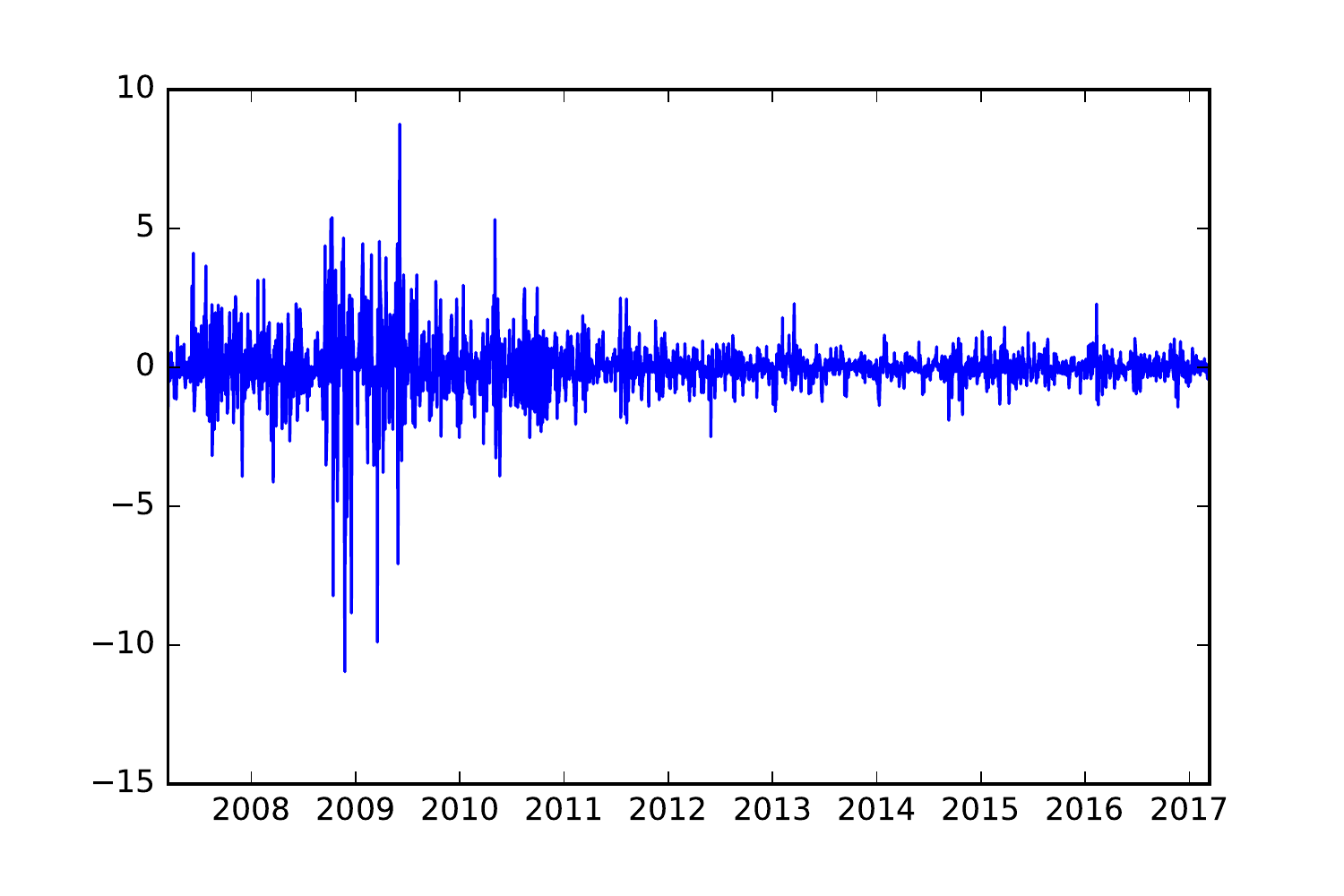}
\includegraphics[width=0.6\textwidth]{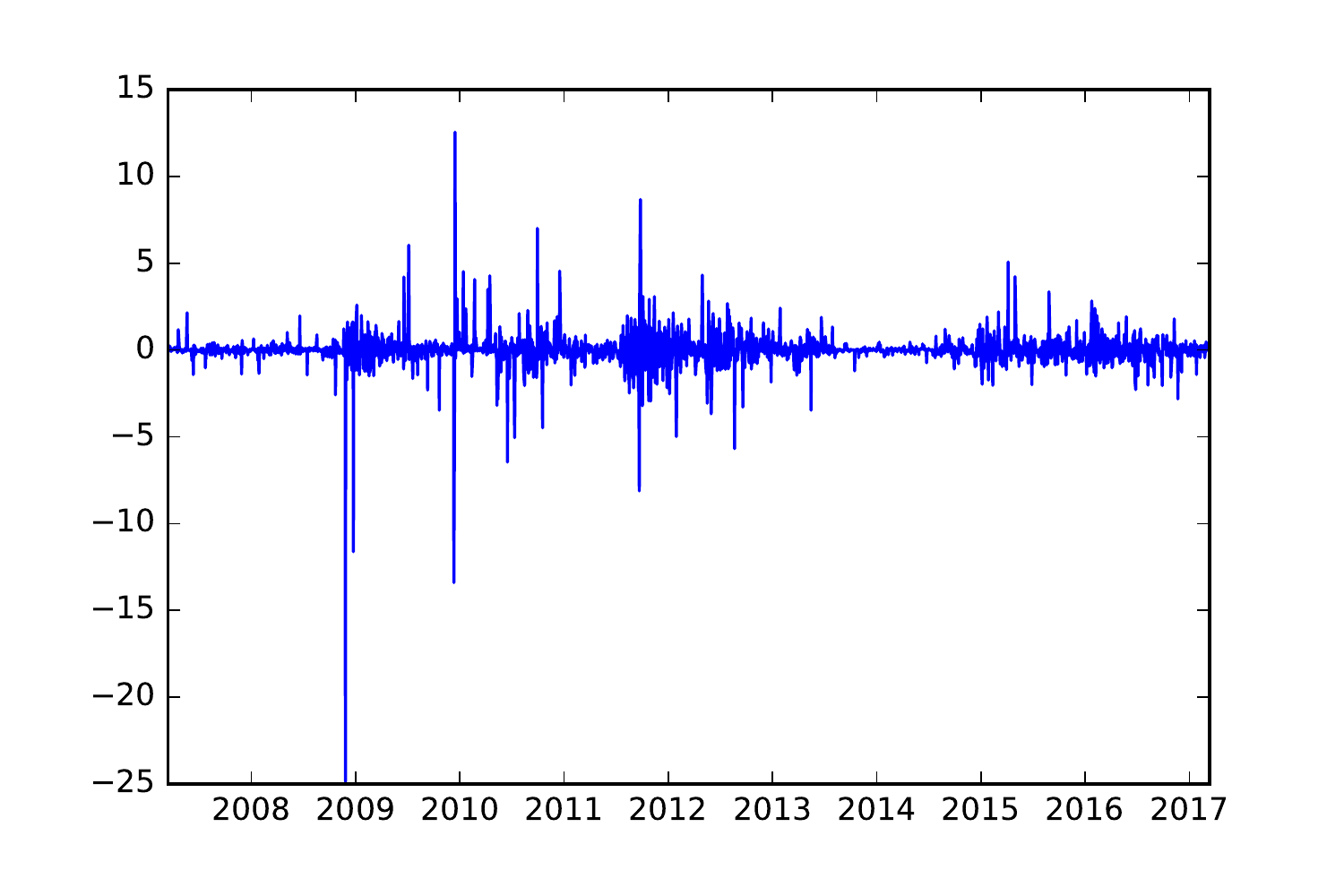}
\includegraphics[width=0.6\textwidth]{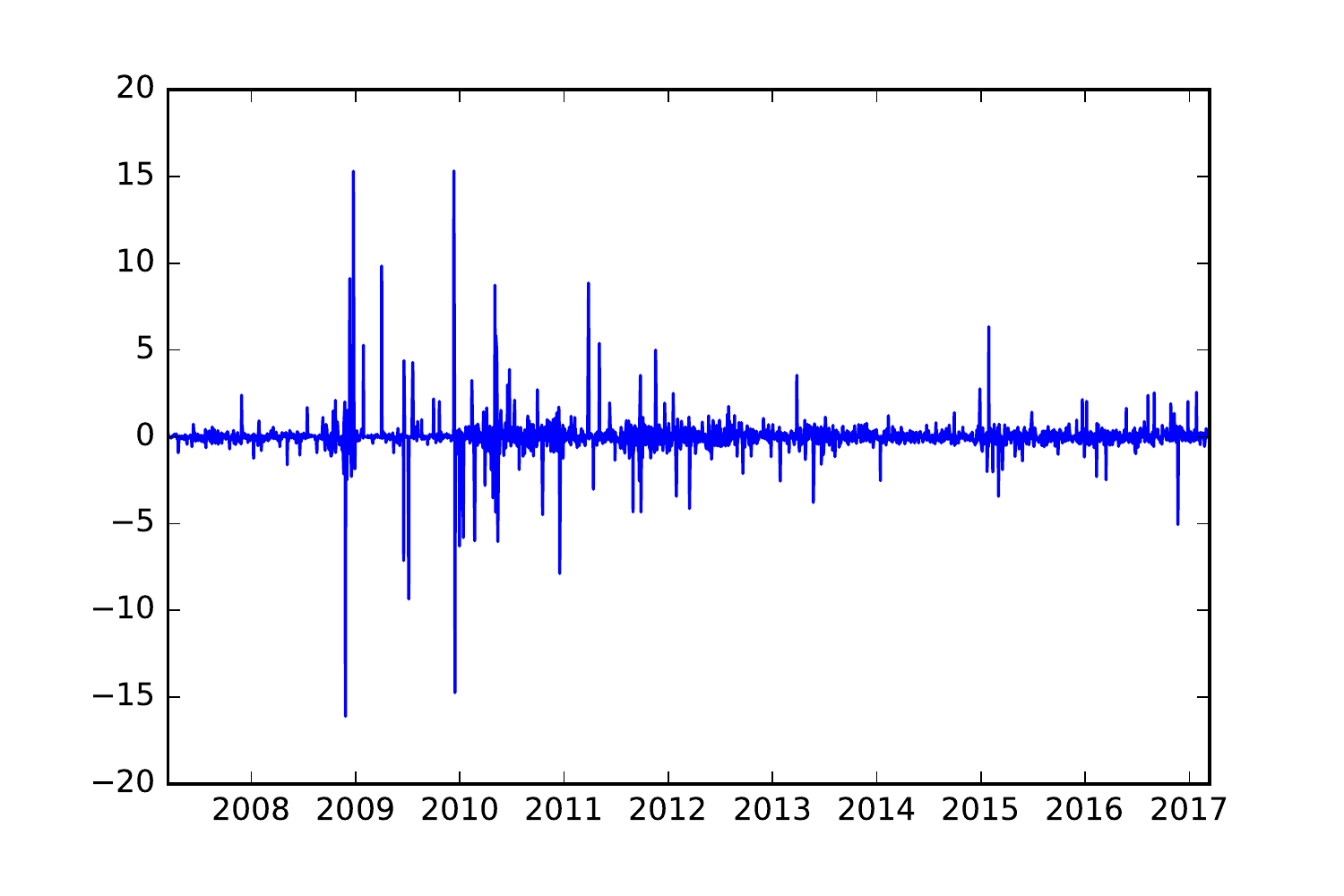}
\caption{Projections ($\xi_i$) of moneyness-index smile log-return on first three components. expiry=10Y, tenor=10Y, currency=USD. }
\label{fig:ts_proj}
\end{figure}

\begin{table}[htbp]
  \centering
  \caption{Correlations between $\{\xi_i\}_i$ for Karhunen-Loève decomposition of moneyness-indexed smile log-return. expiry=10Y, tenor=10Y, currency=USD}
    \begin{tabular}{|l|r|r|r|}
    \toprule
          & \multicolumn{3}{c|}{combinations of projections $\{\xi_i\}_i$} \\
    \midrule
          & \multicolumn{1}{c|}{1st and 2nd} & \multicolumn{1}{c|}{1st and 3rd} & \multicolumn{1}{c|}{2nd and 3rd} \\
    \midrule
    Pearson's correlation & 0.00001\% & 0.00016\% & -0.00007\% \\
    \midrule
    p-value & 99.99975\% & 99.99346\% & 99.99715\% \\
    \bottomrule
    \end{tabular}
  \label{tab:corr_proj}
\end{table}

We shall focus on moneyness-indexed smiles in the following sections for studying the implications on risk management. The projections of their log-returns on the first three components in Figure~\ref{fig:eigen_moneyness_indexed_smile} are shown in Figure~\ref{fig:ts_proj}. These are the times series of $\xi_i$ defined in Equation~\ref{eqn:decomposition_2}. Table~\ref{tab:corr_proj} shows that they are in fact mutually uncorrelated. 


\section{Filtered Historical Simulation for Projected Time Series}
\label{sec:FHS}
Value at Risk (VaR), originally an internal risk management measure of JPMorgan, has become the benchmark of the banking system for evaluating market risk. Despite the critics of not providing the expected shortfall (\cite{dowd1998beyond}) or not being subadditive (\cite{artzner1999coherent}), VaR is an intuitive and easy-to-calculate measure summarizing important risk informations. Market risk comes from different risk factors and the risk factor that we are here concerned with is volatility. But implied volatility is a smile, a surface or in the case of swaption, a cube, which renders the study of its dynamics difficult. As we have shown in the previous sections, Karhunen-Loève decomposition can be adopted to have a concise and accurate description of the dynamics. At most three eigenvectors and the projections on these eigenvectors are needed to characterize the dynamics. Hence the volatility risk factor can be resumed to three mutually uncorrelated process which can be studied separately to evaluate the VaR.\\

The most widely used approach for computing VaR is historical simulation. For a given time series $\{r_t\}_t$ (for example $\{\xi_i(t)\}_t$), we want to calculate the VaR($\alpha$), i.e. the $\alpha$-th quantile of the distribution $F$ of the random process. Unlike a parametric approach which assumes a priori a parametric distribution $F$ and estimates the distribution parameters, historical simulation uses directly the empirical distribution based on a rolling window. For example, for calculating VaR($\alpha$) at time $t$, we generate first of all the empirical cumulative distribution function $\hat{F}$ by using $L$ most recent historical observations $r_{t-L}, r_{t-L+1}, \cdots, r_{t-1}$, where $L$ is the length of the rolling window. Then $VaR_t(\alpha)=\hat{F}^{-1}(\alpha)$. Interpolations are often needed since $\hat{F}$ is a step function.\\

Yet as \cite{cont2001empirical} points out, financial return time series often exhibits volatility clustering phenomenon, i.e. conditional heteroscedasticity. For a more general time series, conditional mean structure is also very common, for example in an autoregressive model. It should be noted that both conditional mean structure and conditional heteroscedasticity structure are compatible with the stationarity of the time series. The later one is usually assumed for time series under study, except for special cases where a unit root can be detected. Despite this compatibility, the existence of conditional mean or conditional heteroscedasticity would render a Historical Simulation less credible, since even though the observations $r_{t-L}, r_{t-L+1}, \cdots, r_{t-1}$ have the same distribution in the \textit{unconditional} sense, they come from \textit{conditionally} different distributions. For example, if conditional volatility at time $t-1$ is larger than that at time $t-L$, then using them simultaneously for estimating $\hat{F}_t$ is not quite reasonable. This is the reason why Filtered Historical Simulation has been proposed.\\

\begin{figure}[htbp]
\centering
\includegraphics[width=0.49\textwidth]{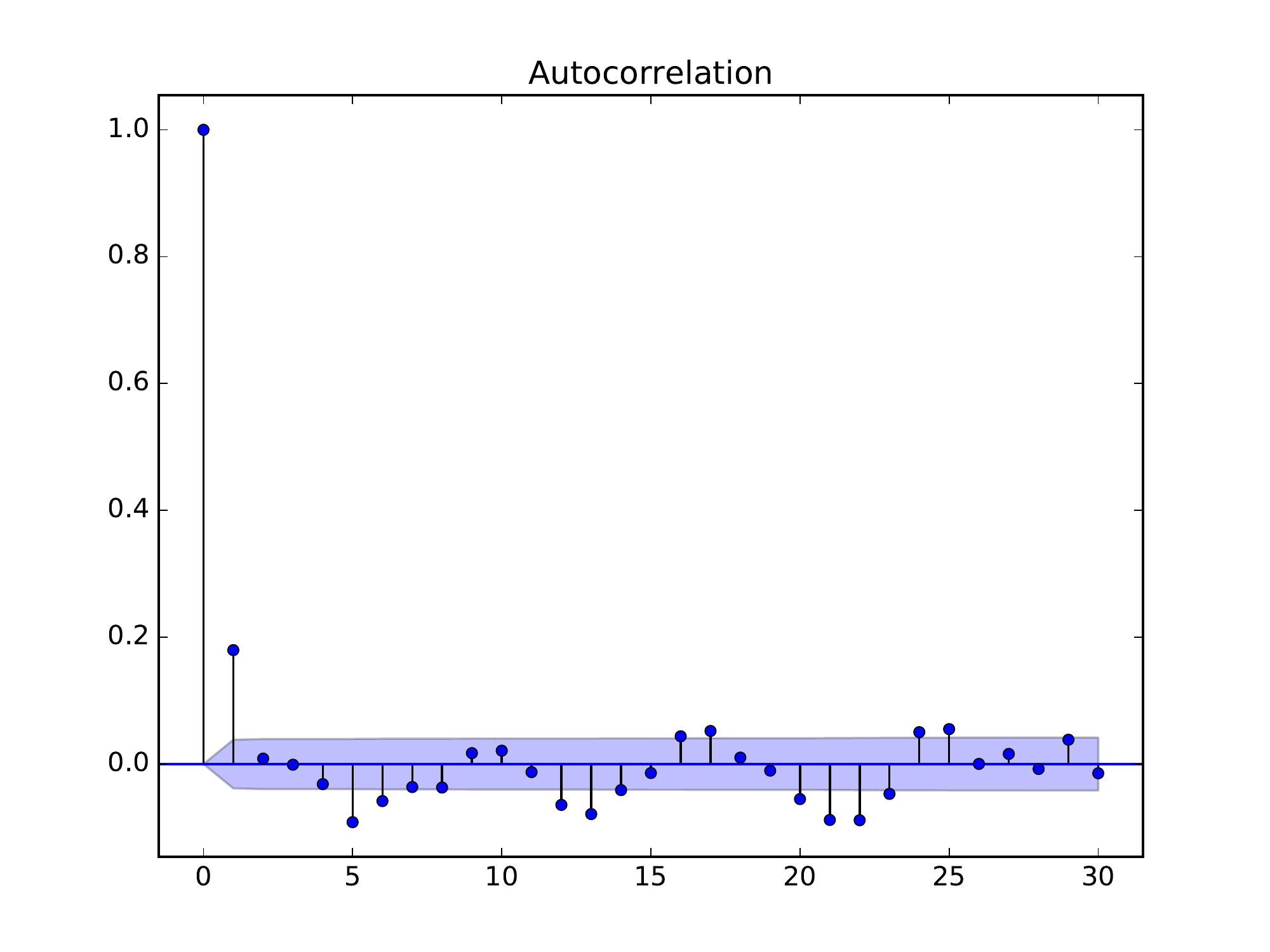}
\includegraphics[width=0.49\textwidth]{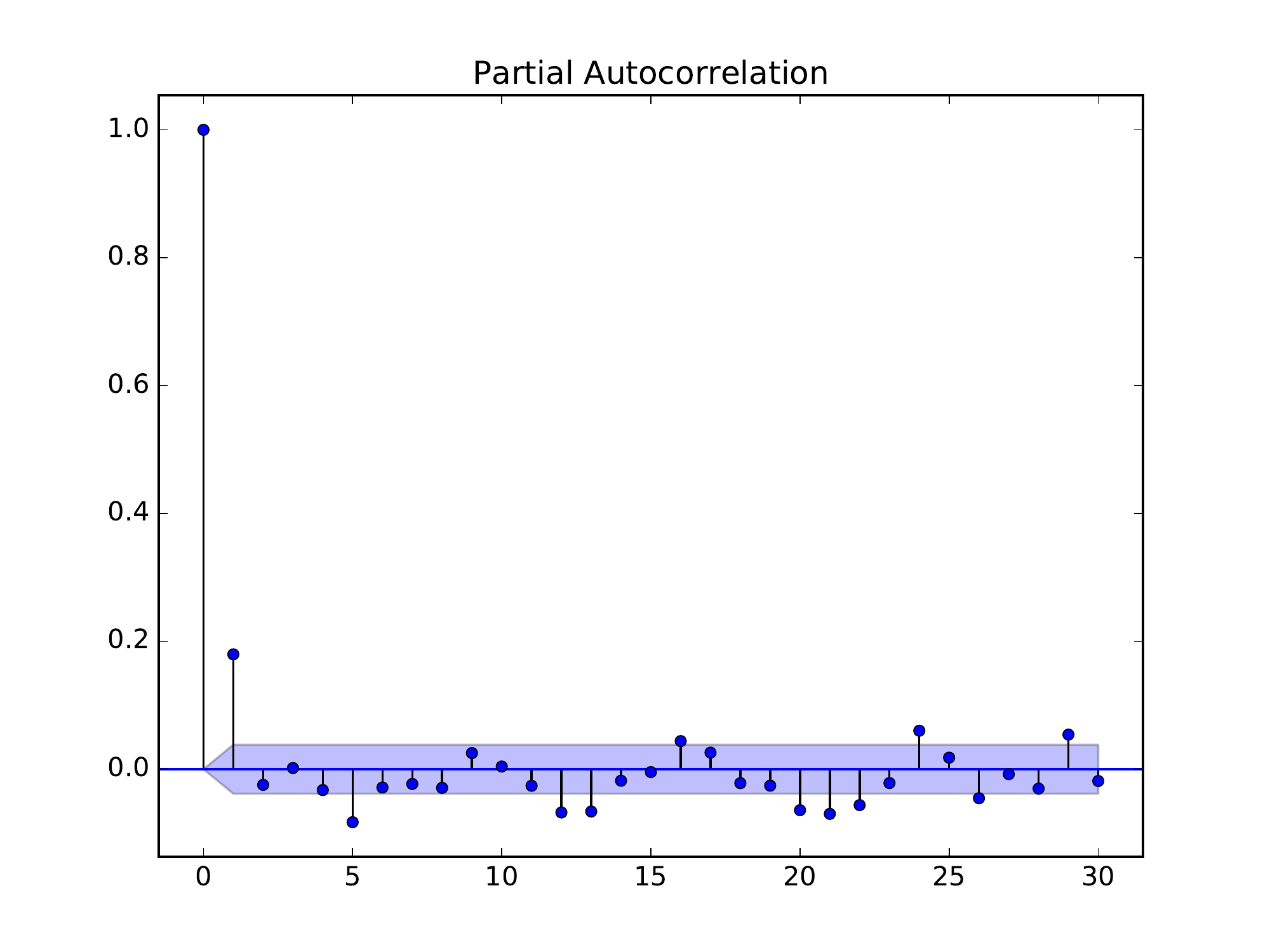}
\caption{ACF and PACF of $\xi_1$, obtained from the Karhunen-Loève decomposition of moneyness-indexed smile.}
\label{fig:acf_pacf_xi1}
\end{figure}

Let's first of all consider the conditional mean structure. For $\xi_1$, whose time series is shown in the first panel of Figure~\ref{fig:ts_proj}, its autocorrelation function (ACF) and partial autocorrelation function (PACF) are shown in Figure~\ref{fig:acf_pacf_xi1}. Looking at the PACF, we can model $\xi_1$ by AR(1), because of the following theorem proved in \cite{roueff2016time}.\\

\begin{theorem}
Let $X$ be a centered weakly stationary process with partial autocorrelation function $\kappa$. Then $X$ is an AR($p$) process if and only if $\kappa(m) = 0$ for all $m > p$.
\end{theorem}

Fitting $\xi_1$ with AR(1), we get the following model:

\begin{equation}
\xi_1(t)=\beta_1\xi_1(t-1)+\epsilon_1(t) \quad \text{with} \quad \beta_1=0.179634
\label{eqn:ar1}
\end{equation}

\begin{figure}[htbp]
\centering
\includegraphics[width=0.8\textwidth]{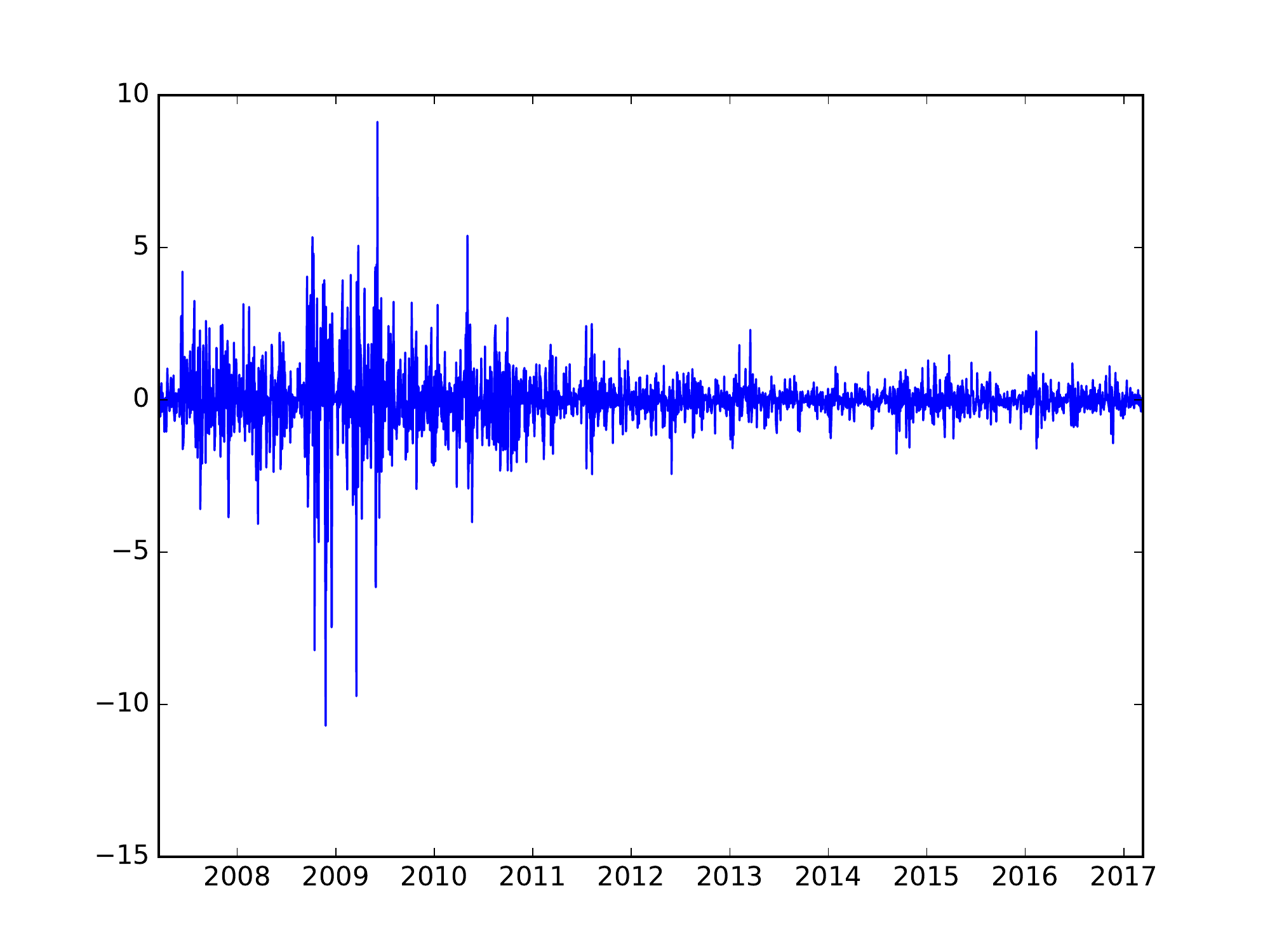}
\includegraphics[width=0.49\textwidth]{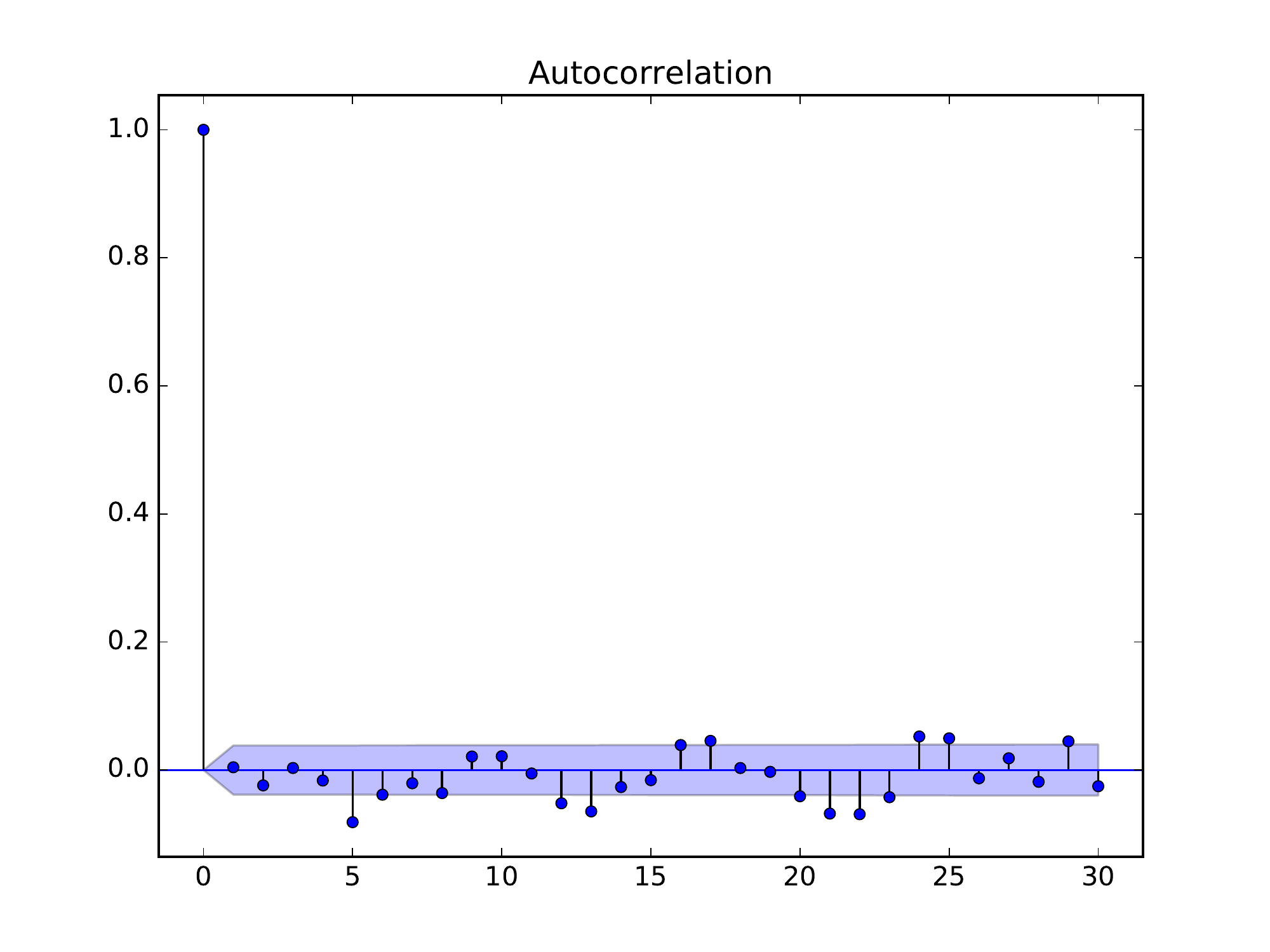}
\includegraphics[width=0.49\textwidth]{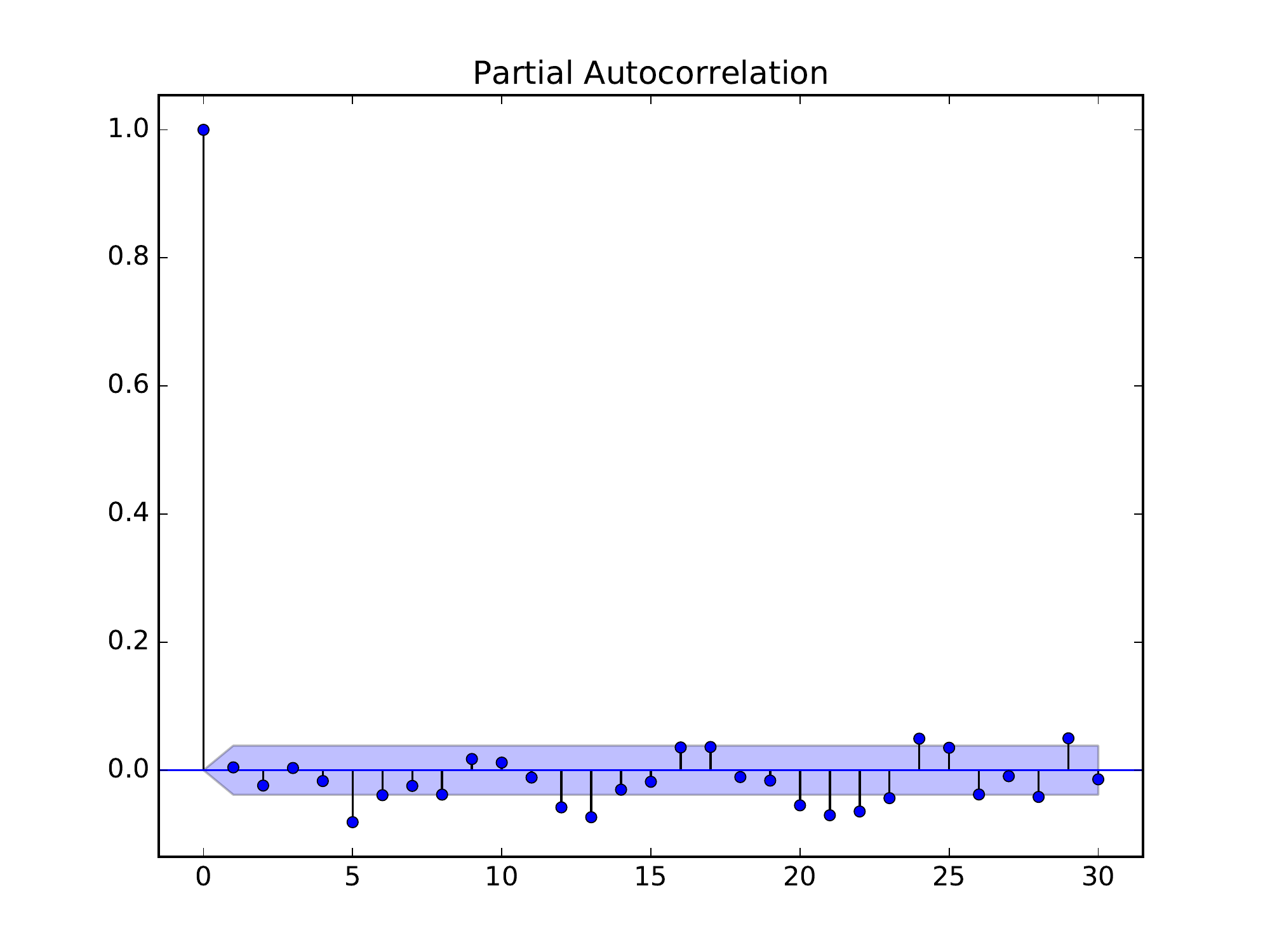}
\caption{Time series, ACF and PACF of $\epsilon_1$, which is obtained from the AR(1) model in Equation~\ref{eqn:ar1} .}
\label{fig:ts_acf_pacf_residue_xi1}
\end{figure}

\begin{figure}[htbp]
\centering
\includegraphics[width=0.49\textwidth]{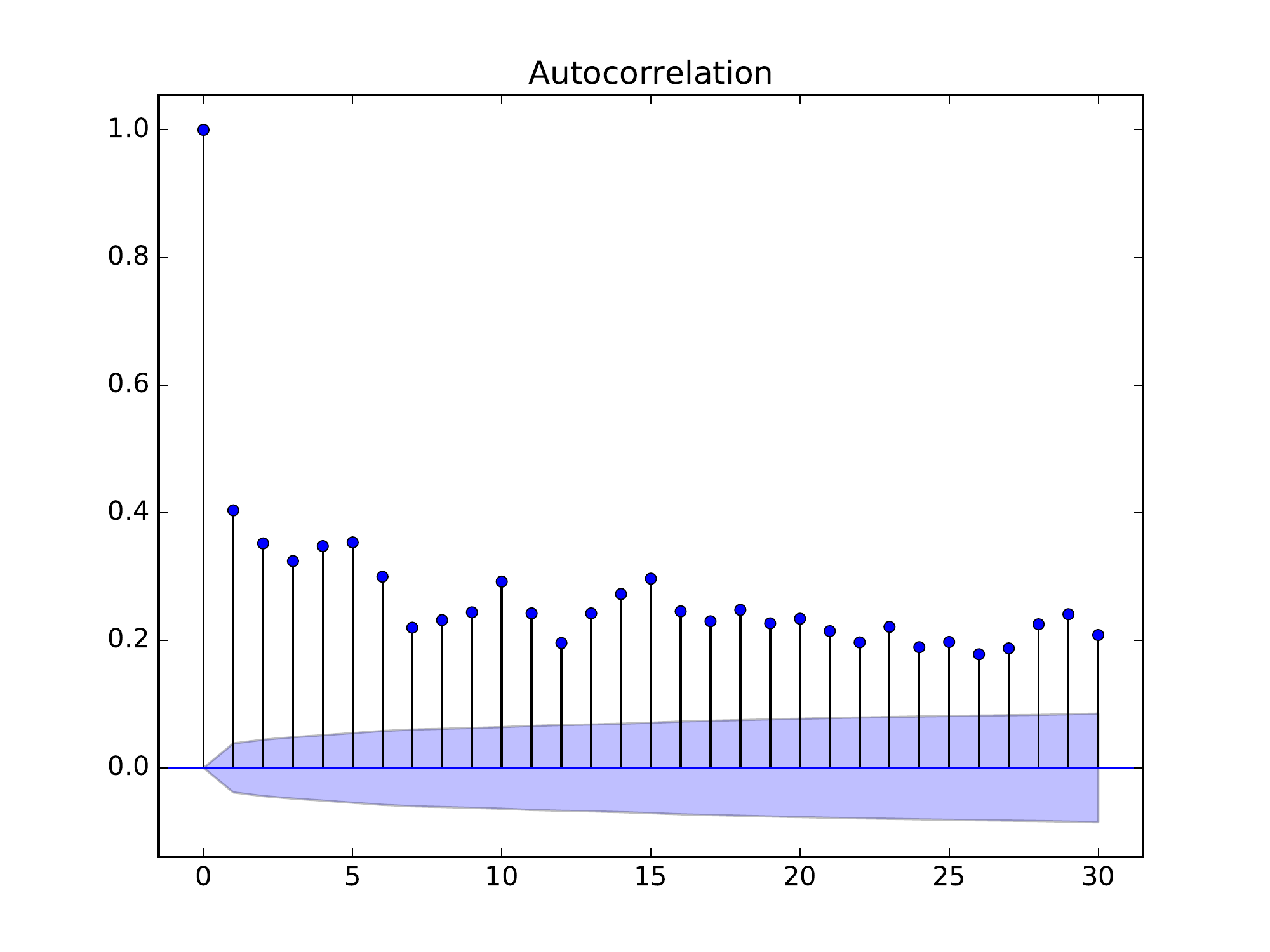}
\includegraphics[width=0.49\textwidth]{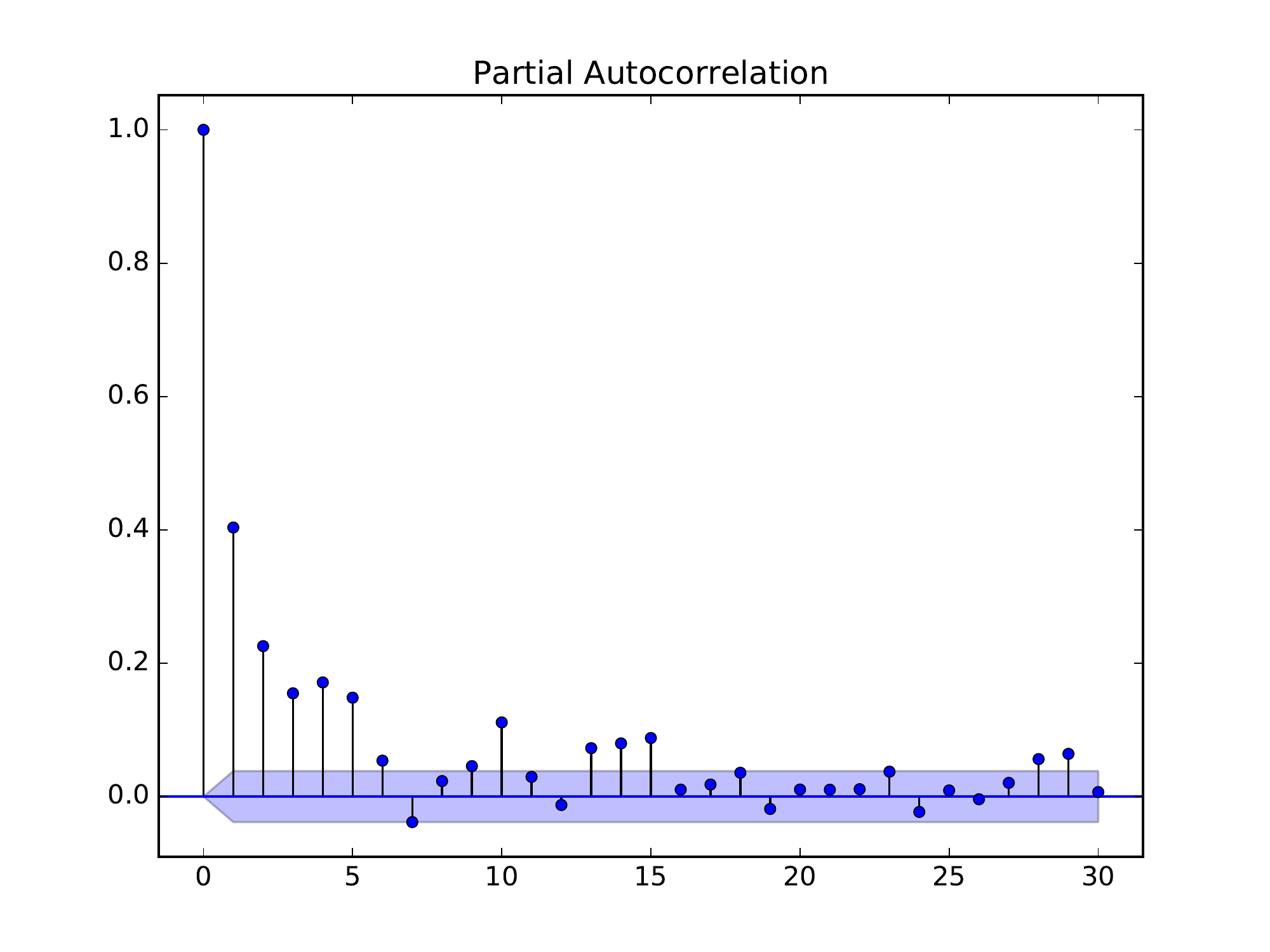}
\caption{ACF and PACF of $|\epsilon_1|$.}
\label{fig:acf_pacf_abs_residue_xi1}
\end{figure}

The time series, as well as ACF and PACF of residues $\{\epsilon_1(t)\}_t$ are shown in Figure~\ref{fig:ts_acf_pacf_residue_xi1}. While the ACF and PACF resemble those of a white noise, the time series clearly have a volatility clustering effect. In fact, taking absolute value of the residues, and recalculating ACF and PACF, we get Figure~\ref{fig:acf_pacf_abs_residue_xi1}. Thus $\epsilon_1$ can be modeled as a weakly white noise but not an IID one. It exhibits conditional heteroscedasticity structure which should be removed before estimating VaR.\\

\begin{figure}[htbp]
\centering
\includegraphics[width=0.8\textwidth]{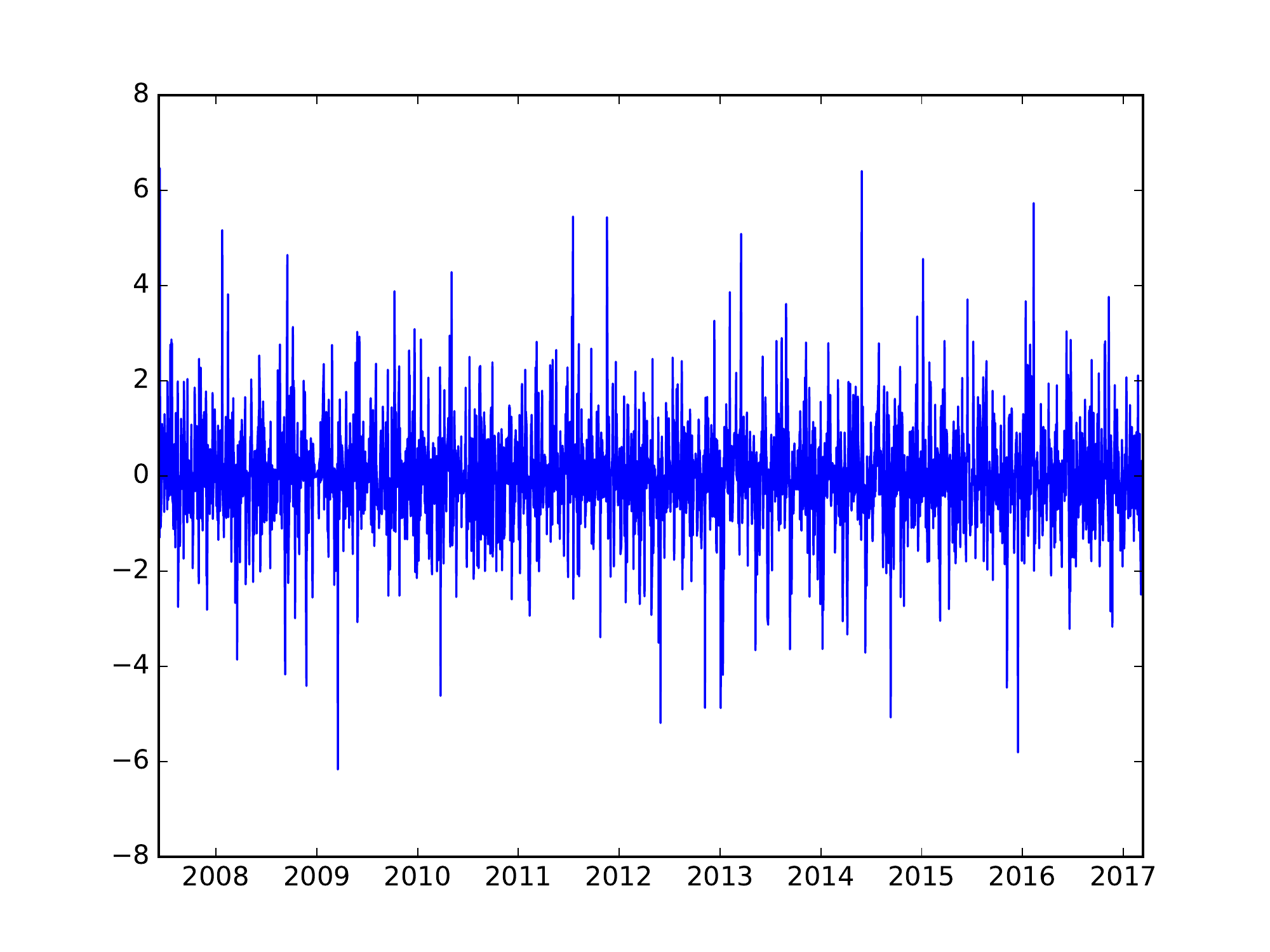}
\caption{Time series of $\epsilon_1/\sigma_1$, where $\sigma_1$ is EWMA volatility of $\epsilon_1$ with decay factor $\theta=0.9$ and volatility estimation rolling window $W=60$.}
\label{fig:ts_descaled_residue_xi1}
\end{figure}

\begin{figure}[htbp]
\centering
\includegraphics[width=0.49\textwidth]{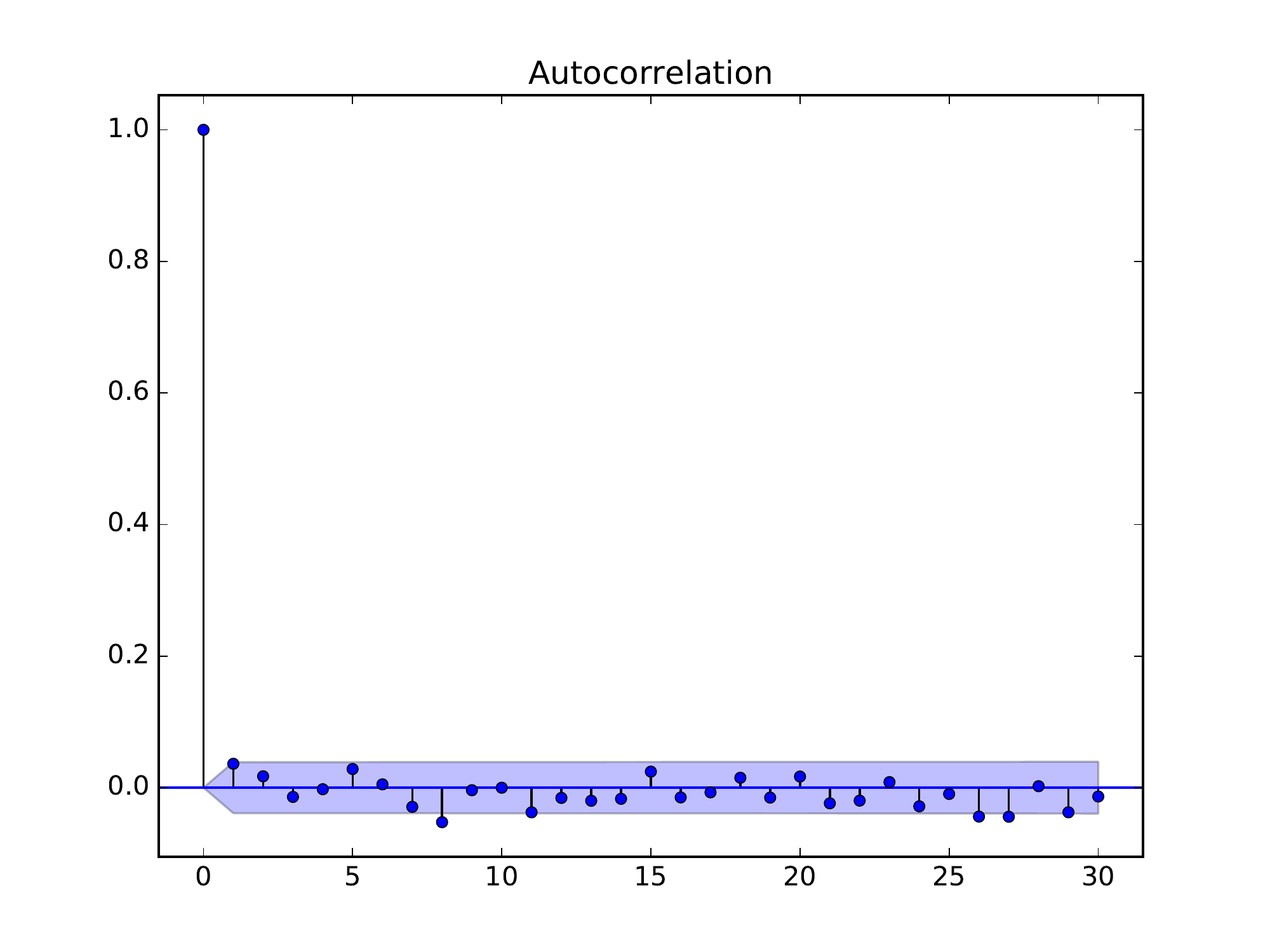}
\includegraphics[width=0.49\textwidth]{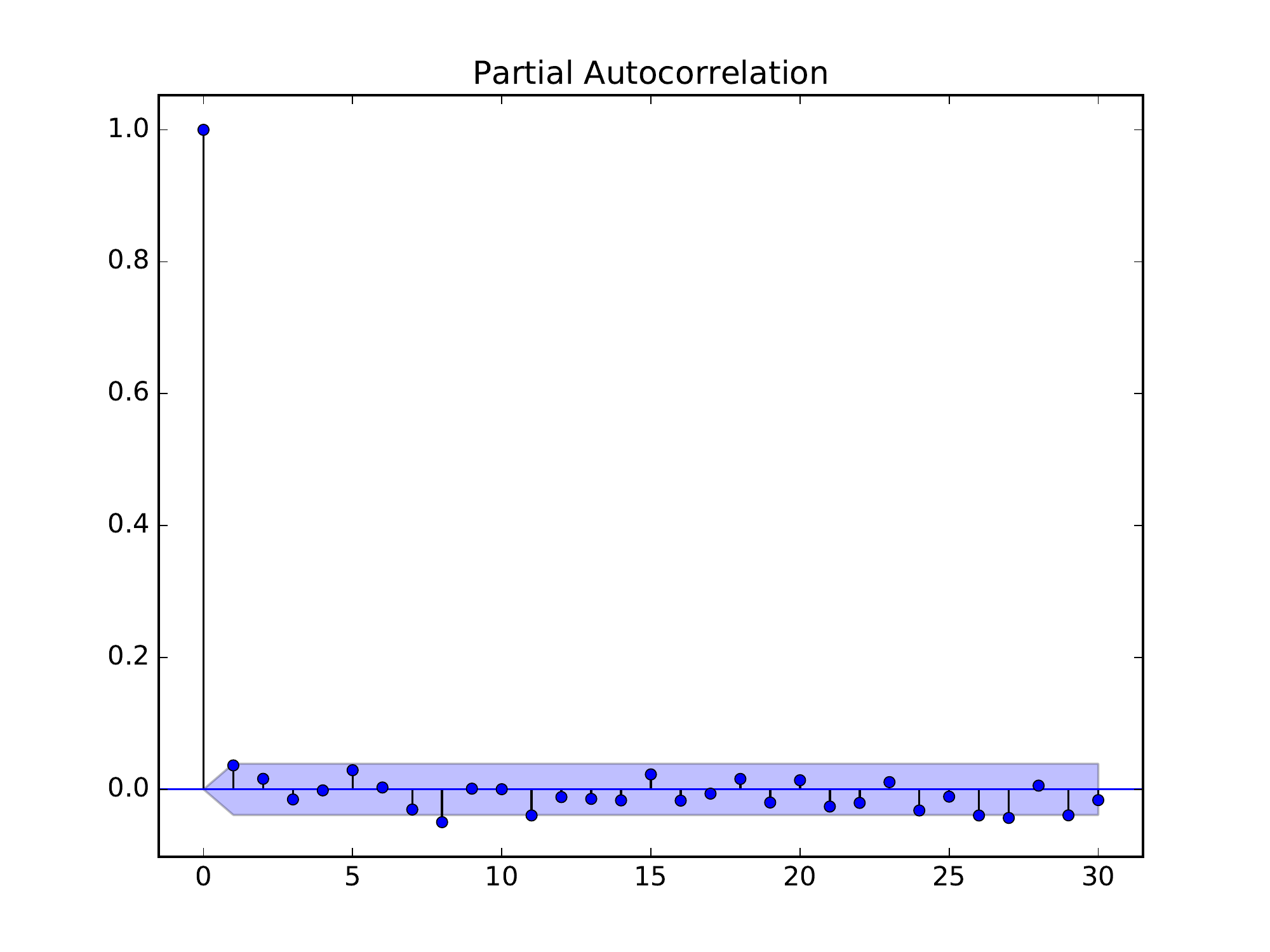}
\caption{ACF and PACF of $|\epsilon_1/\sigma_1|$.}
\label{fig:acf_pacf_abs_descaled_residue_xi1}
\end{figure}

How to estimate conditional volatilities of $\epsilon_1$? The simplest approach would be the standard deviation based on a rolling window, which unfortunately works very badly. More complex considerations such as GARCH($p,q$) model can surely accomplish this task. Here we have adopted a parsimonious approach called Exponentially Weighted Moving Average (EWMA). More specifically, for a centered process $X$, we can estimated its conditional volatility by using the following recursive formula:

\begin{equation}
\sigma^2(t)=\theta \sigma^2(t-1)+(1-\theta)X^2(t-1)
\label{eqn:ewma1}
\end{equation}

where $\theta\in (0,1)$. The method is called ``exponential'' because Equation~\ref{eqn:ewma1} can be applied recursively to get the following equation:

\begin{equation}
\sigma^2(t)=(1-\theta)\sum_{i=1}^W \theta^{i-1}X^2(t-i)+\theta^W\sigma^2(t-W)
\end{equation}

where $W$ is the window length for estimating conditional volatilities. In our study, $W=60$, $\theta=0.9$ so $\theta^W\sigma^2(t-W)$ can be neglected. Calculating EWMA volatilities $\sigma_1$ of residues $\epsilon_1$ and devolatising $\epsilon_1$ by $\sigma_1$, we get the times series $\{\frac{\epsilon_1(t)}{\sigma_1(t)}\}_t$ illustrated in Figure~\ref{fig:ts_descaled_residue_xi1}. The ACF and PACF of $|\frac{\epsilon_1}{\sigma_1}|$ are shown in Figure~\ref{fig:acf_pacf_abs_descaled_residue_xi1}. The volatility clustering phenomenon has been completely removed!\\

In summary, we have firstly removed the conditional mean structure by applying AR(1) on projected time series $\xi_1$ to get the residues $\epsilon_1$, then the conditional heteroscedasticity structure by descaling the residues $\epsilon_1$ by conditional volatilities $\sigma_1$. Now Historical Simulation can be safely applied to $\frac{\epsilon_1}{\sigma_1}$. In our study, the 1st and the 99th rolling quantiles are estimated, which are then ``revolatised'' by multiplying $\sigma_1$ to get the 1st and 99th rolling quantiles of residues $\epsilon_1$. Denote the 1st and 99th quantile respectively $\epsilon_1^{(0.01)}$ and $\epsilon_1^{(0.99)}$, we have:

\begin{equation}
\hat{\epsilon}^{(\alpha)}_1(t)=\sigma_1(t)\hat{F}_{1,t}^{-1}(\alpha)
\end{equation}

where $\sigma_1$ is estimated conditional volatility by using EWMA method, $\hat{F}_{1,t}$ is the rolling empirical cumulative distribution defined by:

\begin{equation}
\hat{F}_{1,t}(x)=\frac{1}{L}\sum_{l=1}^{L}\mathbbm{1}_{x\leq \frac{\epsilon_1(t-l)}{\sigma_1(t-l)}}
\end{equation}

\begin{figure}[htbp]
\centering
\includegraphics[width=\textwidth]{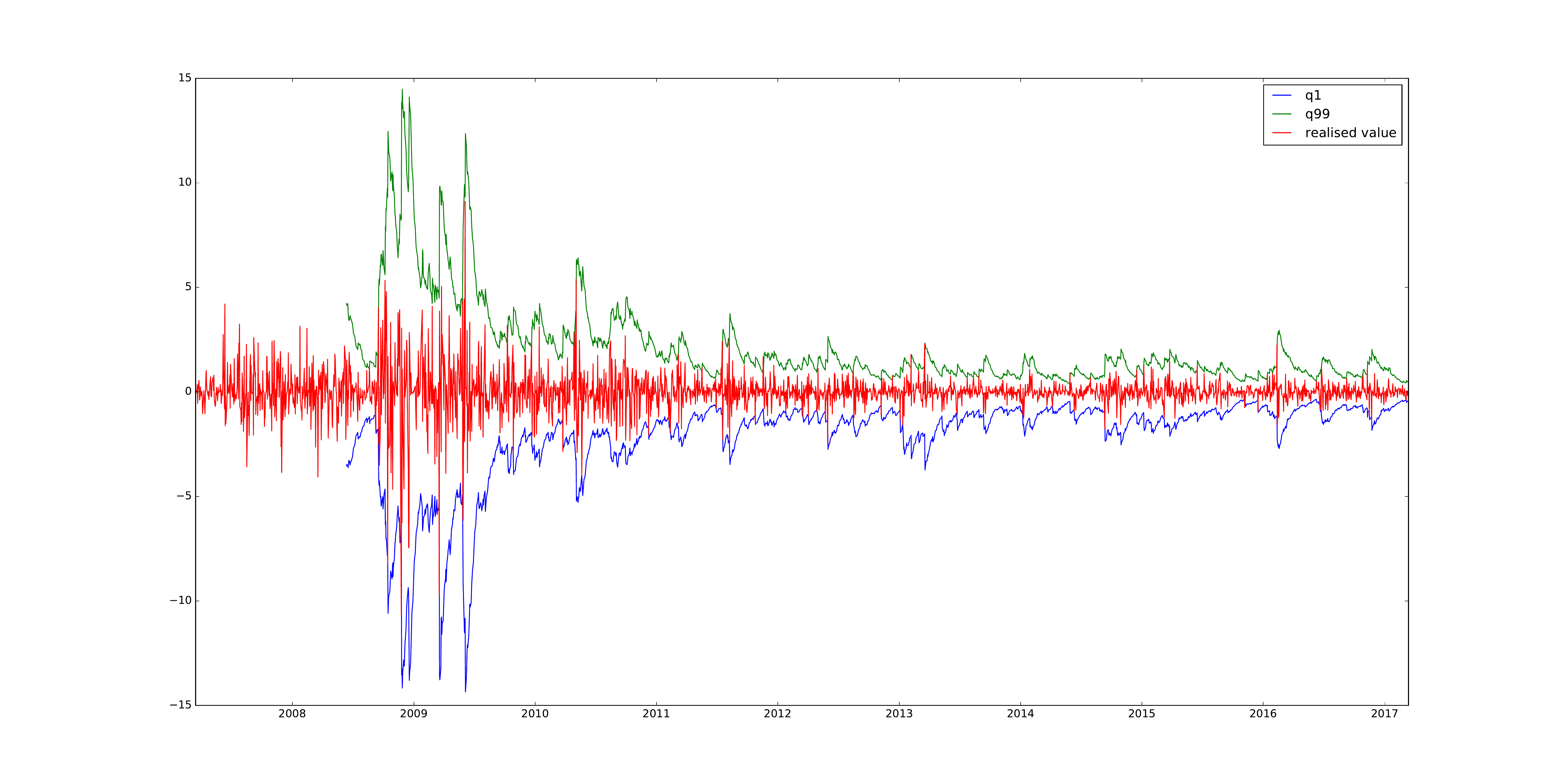}
\caption{Predicted 1st quantile $\hat{\epsilon}_1^{(0.01)}$, 99th quantile $\hat{\epsilon}_1^{(0.99)}$ and the realized time series $\epsilon_1$}
\label{fig:q1_q99_ts_residue}
\end{figure}

The comparison of  $\hat{\epsilon}_1^{(0.01)}$, $\hat{\epsilon}_1^{(0.99)}$ and the realized time series $\epsilon_1$ are illustrated in Figure~\ref{fig:q1_q99_ts_residue}. It can be seen that $\hat{\epsilon}_1^{(0.01)}$ and $\hat{\epsilon}_1^{(0.99)}$ form a good contour of $\epsilon_1$.\\

Furthermore, according to Equation~\ref{eqn:ar1}, the quantiles of $\xi_1$ can be estimated by:

\begin{equation}
\hat{\xi}_1^{(\alpha)}(t)=\beta_1\xi_1(t-1)+\hat{\epsilon}_1^{(\alpha)}(t)
\end{equation}

where $\beta_1=0.179634$, and an extreme parallel shift of smile log-return is $\sqrt{\lambda_1}\hat{\xi}_1^{(\alpha)}(t)e_1(x)$. Consequently, the predicted extreme smile is
\begin{equation}
\hat{I}_1^{(\alpha)}(t,x)=I(t-1, x)+exp(\sqrt{\lambda_1}\hat{\xi}_1^{(\alpha)}(t)e_1(x))
\end{equation}
Note we have used the subscript $1$ to emphasize the fact that the predicted smile is only related to the move along the first principal component. Denote $C(f_t, \kappa, \sigma)$ the pricing function of a payer swaption at time $t$ whose option contract expires at $t+T$. $f_t$ is the forward rate of the fixed leg of the underlying swap contract. Then VaR of the swaption at time $t$ resulting from parallel smile shift would be:
\begin{equation}
C(f_t, \kappa, \hat{I}_1^{(\alpha)}(t, \kappa-f_t))-C(f_{t-1}, \kappa, I(t, \kappa-f_{t-1}))
\end{equation}
What if we want to calculate the VaR resulting from the combined move of the first several principal components? Repeating the procedure described in this section, $\hat{\xi}_2^{(\alpha)}$ and $\hat{\xi}_3^{(\alpha)}$ can be estimated in a similar way\footnote{For some case, the autoregressive structure may not be necessary. This is the case for $\xi_2$ and $\xi_3$ since their PACF look like a white noise.} Hence according to Equation~\ref{eqn:decomposition_2}, the quantile of smile log-return can be calculated by:

\begin{equation}
\hat{u}^{(\alpha)}(t, x)=\sum_{i=1}^3 \sqrt{\lambda_i}\hat{\xi}_i^{(\alpha)}(t)e_i(x)
\label{eqn:var_smile}
\end{equation}

Here we have used only the first 3 eigenvectors as they explain nearly $100\%$ of the dynamics. But there is a caveat here: the VaR of a 3-dimensional vector, which is the case of $\hat{u}^{(\alpha)}(t, x)$, is not well defined. For example, in the case of moneyness-indexed smile, if we take $\alpha=0.01$ and look at Figure~\ref{fig:eigen_moneyness_indexed_smile}, then $\hat{\xi}_1^{(0.01)}(t)e_1(x)$ means an extreme \textit{negative} parallel shift, $\hat{\xi}_2^{(0.01)}(t)e_2(x)$ means an extreme counter-clockwise rotation and $\hat{\xi}_3^{(0.01)}(t)e_3(x)$ means an extreme flattening of the smile. Thus $\hat{u}^{(0.01)}(t, x)$ defined in Equation~\ref{eqn:var_smile} incorporates these three aspects. Yet we might as well define the VaR as the combination of extreme \textit{positive} parallel shift, extreme counter-clockwise rotation and extreme flattening of the smile, i.e. $\hat{\xi}_1^{(0.99)}(t)e_1(x)+\hat{\xi}_2^{(0.01)}(t)e_2(x)+\hat{\xi}_3^{(0.01)}(t)e_3(x)$. The definition of smile VaR would thus be based on the concerns in practice and we shall stick with this definition for the simplicity of notation. Based on Equations~\ref{eqn:return} and \ref{eqn:var_smile}, if we are at time $t-1$, we can estimate the extreme cases of volatility smile (or surface etc.) at time $t$ by:

\begin{equation}
\hat{I}^{(\alpha)}(t,x)=I(t-1, x)+exp(\hat{u}^{(\alpha)}(t, x))
\end{equation}

Consequently the VaR of the swaption at time $t$ resulting from smile move would be:
\begin{equation}
C(f_t, \kappa, \hat{I}^{(\alpha)}(t, \kappa-f_t))-C(f_{t-1}, \kappa, I(t, \kappa-f_{t-1}))
\end{equation}


\section{Do Reconstructed Volatility Smiles Violate Non-Arbitrage Condition?}
\label{sec:arbitrage}
As it is well-known, non-arbitrage condition is the most basic assumption in derivative pricing. It has direct implications on the prices of options, even in model-free conditions. In the case of a payer swaption, whose option contract is a call option, the price of the swaption should be decreasing and convex as a function of the strike. To demonstrate the idea, it suffices to investigate a simple European call option $Call(S_0, \kappa, \sigma)$. It should satisfy the following conditions:

\begin{eqnarray}
\frac{\partial Call(S_0, \kappa, \sigma)}{\partial K} & \leq & 0\\
\frac{\partial^2 Call(S_0, \kappa, \sigma)}{\partial K^2} & \geq & 0
\label{eqn:non_arbitrage}
\end{eqnarray}

The first equation is straightforward if we look at the original pricing function $Call(S_0, \kappa, \sigma)=\mathbb{E}(S_T-\kappa)_+$. For the second one, imagine a portfolio X consisting of a long position of $\lambda$ calls with strike $\kappa_1$, a long position of $1-\lambda$ calls with strike $\kappa_2$, and a short position of a call with strike $\lambda \kappa_1+(1-\lambda)\kappa_2$. According to the convexity of the function $x_+$ and using Jensen's inequality, we have:

\begin{equation}
\lambda (S_T-\kappa_1)_+ + (1-\lambda)(S_T-\kappa_2)_+ -(S_T-\lambda \kappa_1-(1-\lambda)\kappa_2)_+ \geq 0
\end{equation}

This means that the terminal value of the portfolio at maturity $T$ is non-negative. Hence taking the expectation, the price of the portfolio at time $0$ should also be non-negative. Because $\lambda$ can be arbitrary, we get the convexity of $Call$ with respect to $\kappa$.\\

\begin{figure}[htbp]
\centering
\includegraphics[width=0.9\textwidth]{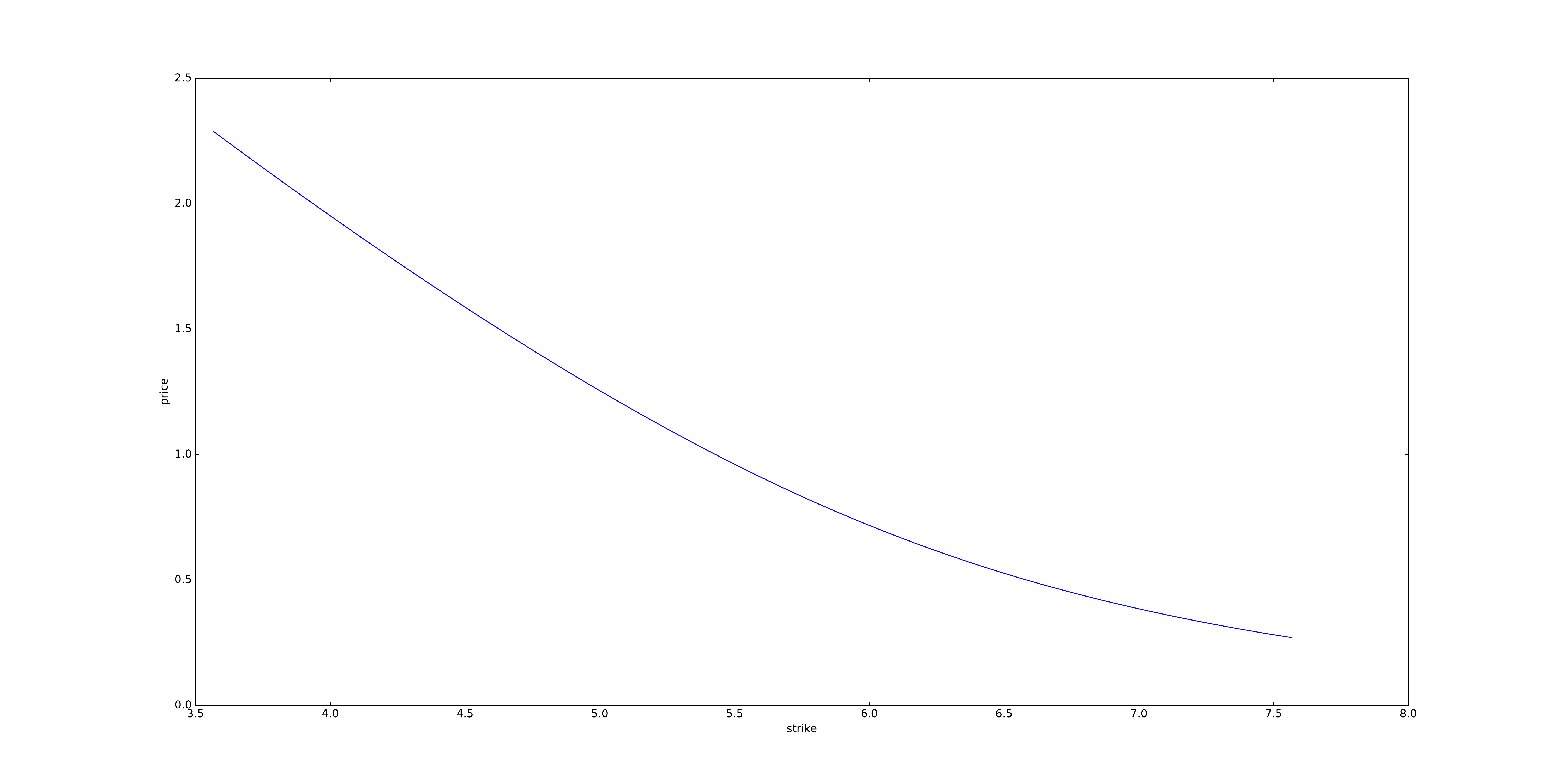}
\caption{An example of pricing function with respect to strike for a payer swaption, when extreme predicted implied volatilities are used. t=2008-06-12, expiry T=10Y, tenor=10Y, currency: USD, $\alpha=0.01$.}
\label{fig:example_convexity}
\end{figure}

Does $\hat{I}^{(\alpha)}(t,x)$ constructed in the previous section violate the non-arbitrage condition? We should resort to the price function in order to answer to this question. Using the notations from previous sections, the predicted extreme price of the swaption would be $C(f_t, \kappa, \hat{I}^{(\alpha)}(t, \kappa-f_t))$. Note that we have used $\kappa-f_t$ to replace $x$ because $x$ represents moneyness defined in \ref{eqn:moneyness}. An example of the price function with respect to strike is shown in Figure~\ref{fig:example_convexity} which observes conditions in Equations~\ref{eqn:non_arbitrage}. In fact, for the historical period that we have studied (2007-2017), none of the days violates the non-arbitrage condition. This shows that the Karhunen-Loève decomposition, together with the Filtered Historical Simulation that we have adopted, is well compatible with the pricing framework.

\section{Backtesting VaR}
\label{sec:backtest}

Is the VaR of $\epsilon_1$ (or of $\xi_1$) well estimated?  From Figure~\ref{fig:q1_q99_ts_residue} we see that $\hat{\epsilon}_1^{(0.01)}$ and $\hat{\epsilon}_1^{(0.99)}$ form a good contour of $\epsilon_1$, but what is exactly the ``goodness'' of a VaR estimation? \cite{campbell2006review} summarizes backtesting approaches for VaR. In order to proceed with the backtesting, we should first of all define ``hit sequences'' as follows:

\begin{equation}
h^{(0.01)}(t) = \left \{
\begin{array}{rl}
1 & \text{if } \epsilon_1(t)\leq \hat{\epsilon}_1^{(0.01)}(t)\\
0 & \text{if } \epsilon_1(t)> \hat{\epsilon}_1^{(0.01)}(t)
\end{array}\right .
\end{equation}

\begin{equation}
h^{(0.99)}(t) = \left \{
\begin{array}{rl}
1 & \text{if } \epsilon_1(t)\geq \hat{\epsilon}_1^{(0.99)}(t)\\
0 & \text{if } \epsilon_1(t)< \hat{\epsilon}_1^{(0.99)}(t)
\end{array}\right .
\end{equation}

so that the hit sequence tallies the history of whether or not a move in excess of the reported VaR has been realized.\\

\cite{christopherson1998conditioning} points out that the problem of determining the accuracy of a VaR model can be reduced to the problem of determining whether the hit sequence, $h^{(\alpha)}(t)$, satisfies the following two properties:

\begin{enumerate}
\item Unconditional Coverage Property: The probability of realizing a move in excess of the reported VaR must be precisely $\alpha\times 100\%$ ($(1-\alpha)\times 100\%$) for $\alpha<0.5$ ($\alpha>0.5$), i.e. $\mathbb{P}(h^{(\alpha)}(t)=1)=\alpha$ ($\mathbb{P}(h^{(\alpha)}(t)=1)=1-\alpha$). Otherwise, we have either overestimated or underestimated VaR.\\

To test this property, \cite{kupiec1995techniques} constructed the following test statistic $POF$ (proportion of failures) for $\alpha<0.5$. The $POF$ for $\alpha>0.5$ can be defined analogously. Under null hypothesis:

\begin{equation}
H_0: \hat{\alpha}=\alpha,
\end{equation}

$POF$ asymptotically follows $\chi^2_1$, i.e. chi-square distribution with one degree of freedom.

\begin{equation}
POF=-2\text{ln}(\frac{(1-\alpha)^{T_0}\alpha^{T_1}}{(1-\hat{\alpha})^{T_0}\hat{\alpha}^{T_1}})
\end{equation}

where

\begin{equation}
\hat{\alpha}=\frac{T_1}{T_0+T_1}, \quad T_1=\sum_{t=1}^{T}h^{(\alpha)}(t), \quad T_0=T-T_1
\end{equation}

\item Independence Property: Any two elements of the hit sequence, $(h^{(\alpha)}(t+j), h^{(\alpha)}(t+k))$ must be independent from each other. Intuitively, this condition requires that the previous history of VaR violations, $\{\dots, h^{(\alpha)}(t-1), h^{(\alpha)}(t)\}$, must not convey any information about whether or not an additional VaR violation, $h^{(\alpha)}(t+1)$, will occur. If not, previous VaR violations presage a future VaR violation, which further suggests a lack of responsiveness in the reported VaR measure as changing market risks fail to be fully incorporated into the reported VaR.\\

\cite{christopherson1998conditioning} proposed the following test statistic $IND$ for $\alpha<0.5$. The test statistic for $\alpha>0.5$ can be defined analogously. Under null hypothesis:

\begin{equation}
H_0: \quad \hat{\alpha}_{01}=\hat{\alpha}_{11}=\bar{\alpha}
\end{equation}

$IND$ asymptotically follows $\chi^2_1$.

\begin{equation}
IND=-2\text{ln}(\frac{(1-\bar{\alpha})^{T_{00}+T_{10}}\bar{\alpha}^{T_{01}+T_{11}}}{(1-\hat{\alpha}_{01})^{T_{00}}\hat{\alpha}^{T_{01}}_{01}(1-\hat{\alpha}_{11})^{T_{10}}\hat{\alpha}^{T_{11}}_{11}})
\end{equation}

where
\begin{equation}
\hat{\alpha}_{ij}=\frac{T_{ij}}{T_{i0}+T_{i1}}, \quad \bar{\alpha}=\frac{T_{01}+T_{11}}{T_{00}+T_{01}+T_{10}+T_{11}}
\end{equation}

and $T_{ij}$ is the frequency of a $i$ value followed by a $j$ value in the hit sequence. 

\end{enumerate}

These two properties of the ``hit sequence'' $\{h^{(\alpha)}(t)\}_{t=1}^T$ are often combined into the single statement:

\begin{equation}
h^{(\alpha)}(t)\overset{i.i.d.}{\sim}\left \{
\begin{array}{ll}
Bernoulli(\alpha) & \text{if } \alpha<0.5\\
Bernoulli(1-\alpha) & \text{if } \alpha>0.5
\end{array}\right .
\end{equation}

Now we can use these these two tests for $\epsilon_1$. The results are shown in Table~\ref{tab:backtest} and all tests have been successfully passed, which suggests that the estimated VaR well corresponds to the unconditional probability and adjusts sufficiently fast to incorporate conditional information. Note that we can also conduct these tests for $\xi_1$, which would yield similar results.

\begin{table}[htbp]
  \centering
  \caption{Backtesting results for $\epsilon_1$}
    \begin{tabular}{|c|r|r|r|r|r|r|}
    \toprule
          & \multicolumn{1}{c|}{Kupiec test} & \multicolumn{5}{c|}{Christofferson test} \\
    \midrule
          & \multicolumn{1}{c|}{p-value} & \multicolumn{1}{c|}{$T_{00}$} & \multicolumn{1}{c|}{$T_{01}$} & \multicolumn{1}{c|}{$T_{10}$} & \multicolumn{1}{c|}{$T_{11}$} & \multicolumn{1}{c|}{p-value} \\
    \midrule
    q1    & 29.36\% & 2226  & 28    & 28    & 0     & 40.42\% \\
    \midrule
    q99   & 65.30\% & 2233  & 24    & 24    & 1     & 27.71\% \\
    \bottomrule
    \end{tabular}%
  \label{tab:backtest}%
\end{table}%


\section{Conclusion}
\label{sec:conclusion}

Managing market risk has always been crucial for ensuring the soundness and stability of financial institutions. Among all kinds of risk factors, volatility risk is one of the most important and has been the central concern of this article. More specifically, we are concerned with volatilities of swaptions which are one of the most liquid interest rate derivatives. Instead of using a local volatility or a stochastic volatility model, which gives an infinitesimal description of the volatility dynamics, we have adopted Bachelier model together with implied volatilities. This relatively simple approach is appropriate because of the liquid characteristic of swaptions, and the advantage of being observable in the markets. The dynamics of volatility smiles or surfaces is studied by Karhunen-Loève decomposition, a generalized version of Principal Component Analysis conceived for functions. The decomposition gives a concise and precise description of the dynamics, using at most 3 principal components. As swaption has three important parameters (moneyness, expiry, tenor) for indexing the volatility smile, the decomposition has been separately conducted for these 3 cases and different behaviors have been observed along these three dimensions. For moneyness- or tenor-indexed smile, the first three components can generally be explained as parallel shift, rotation and convexity change. While for expiry-indexed smile, the interpretation is more difficult. What's more, volatilities along different dimensions have inter-connections, which explains the difference between a marginal of a 2-dimensional principal component and the corresponding 1-dimensional principal component.\\

Thanks to null correlations, projections on different principal components can be investigated separately for evaluating VaR. We have particularly paid attention to the conditional mean structure and the conditional heteroscedasticity structure of the time series, in order to have a more ``stable'' time series which is favorable for Historical Simulation. This approach is called Filtered Historical Simulation in the general sense. Extreme moves of volatility smile can thus be estimated in a holistic way and are compatible with non-arbitrage hypothesis. Moreover, estimated VaR well passes Kupiec's unconditional coverage test and Christofferson's independence test, which are the most widely accepted criteria for VaR backtesting. Consequently, volatility risk of a swaption portfolio can be consistently evaluated.\\

It should be noted, nonetheless, that special attention should be paid if we want to extend the method to other financial products. For less liquid products, modeling the dynamics of the underlying using local or stochastic volatility model remains a crucial step for pricing the products and managing market risk. Further investigations should be made in order to appropriately calibrate the parameters and to manage volatility risk.

\bibliography{bib_report}
\bibliographystyle{plainnat}

\end{document}